\documentclass{aa}  

\usepackage{graphicx}
\usepackage{txfonts}

\newcommand{\iras}{IRAS \mbox{18264$-$1152}}
\newcommand{\hmol}{H$_2$}
\newcommand{\brg}{Br$\gamma$}
\newcommand{\feii}{[\ion{Fe}{II}]}
\newcommand{\ee}[1]{\mbox{${} \times 10^{#1}$}} 
\newcommand{\lsun}{\mbox{$\mathrm{L}_\odot$}}
\newcommand{\msun}{\mbox{$\mathrm{M}_\odot$}}
\newcommand{\kms}{\mbox{km\,s$^{-1}$}}

\newcommand{\micron}{\mbox{$\mu$m}}
\newcommand{\av}{$A_\mathrm{V}$}
\newcommand{\kband}{\textit{K}-band}

\begin{document}
   \title{NIR jets from a clustered region of massive star formation}
   \subtitle{Morphology and composition in the IRAS 18264$-$1152 region}
    
   \author{A. R. Costa Silva\inst{1,2,3}
          \and R. Fedriani\inst{1}
          \and J. C. Tan\inst{1,4}
          \and A. Caratti o Garatti\inst{5,6}
          \and S. Ramsay\inst{7}
          \and V. Rosero\inst{8}
          \and G. Cosentino\inst{1}
          \and P. Gorai\inst{1}
          \and S. Leurini\inst{9}
          }

   \institute{
    Department of Space, Earth and Environment, Chalmers University of Technology, SE-412 96 Gothenburg, Sweden\\
    \email{ana.rita@astro.up.pt}\\
    \email{ruben.fedriani@chalmers.se}
        \and
    Instituto de Astrofísica e Ciências do Espaço, CAUP, Rua das Estrelas, 4150-762, Porto, Portugal
        \and
    Departamento de Física e Astronomia, Faculdade de Ciências, Universidade do Porto, Rua do Campo Alegre 687, 4169-007 Porto, Portugal
        \and
    Department of Astronomy, University of Virginia, Charlottesville, Virginia 22904, USA
        \and
    INAF - Osservatorio Astronomico di Capodimonte, via Moiariello 16, 80131, Napoli, Italy
        \and
    Dublin Institute for Advanced Studies, School of Cosmic Physics, Astronomy \& Astrophysics Section, 31 Fitzwilliam Place, Dublin 2, Ireland
        \and
    European Southern Observatory, Karl-Schwarzschild-Strasse 2, 85748, Garching bei München, Germany
        \and
    National Radio Astronomy Observatory, 1003 Lopezville Rd., Socorro, New Mexico 87801, USA
        \and
    INAF - Osservatorio Astronomico di Cagliari, Via della Scienza 5, 09047 Selargius (CA), Italy}

   \date{Received 11 October 2021 / Accepted 3 December 2021}
   
 
  \abstract
   {Massive stars play crucial roles in determining the physical and chemical evolution of galaxies. However they form deeply embedded in their parental clouds, making it challenging to directly observe these stars and their immediate environments. It is known that accretion and ejection processes are intrinsically related, thus observing the massive protostellar outflows can provide crucial information about the processes governing massive star formation very close to the central engine.}
   {We aim to probe the \iras\ (also known as G19.88-0.53) high-mass star-forming complex in the near infrared (NIR) through its molecular hydrogen (\hmol) jets to analyse the morphology and composition of the line emitting regions and to compare with other outflow tracers.}
   {We observed the \hmol\ NIR jets via K-band (1.9\,\micron\,-\,2.5\,\micron) observations obtained with the integral field units VLT/SINFONI and VLT/KMOS. VLT/SINFONI provides the highest NIR angular resolution achieved so far for the central region of \iras\ ($\sim$\,0.2\arcsec). We compared the geometry of the NIR outflows with that of the associated molecular outflow, probed by CO (2$-$1) emission mapped with the Submillimeter Array.}
   {We identify nine point sources in the SINFONI and KMOS fields of view. Four of these display a rising continuum in the K-band and are \brg\ emitters, revealing that they are young, potentially jet-driving sources. The spectro-imaging analysis focusses on the \hmol\ jets, for which we derived visual extinction, temperature, column density, area, and mass. The intensity, velocity, and excitation maps based on \hmol\ emission strongly support the existence of a protostellar cluster in this region, with at least two (and up to four) different large-scale outflows, found through the NIR and radio observations. We compare our results with those found in the literature and find good agreement in the outflow morphology. This multi-wavelength comparison also allows us to derive a stellar density of $\sim$\,4000\,stars\,pc$^{-3}$.}
   {Our study reveals the presence of several outflows driven by young sources from a forming cluster of young, massive stars, demonstrating the utility of such NIR observations for characterising massive star-forming regions. Moreover, the derived stellar number density together with the geometry of the outflows suggest that stars can form in a relatively ordered manner in this cluster.}

   \keywords{ISM: jets and outflows – ISM: kinematics and dynamics – stars: pre-main sequence – stars: massive – stars: individual: IRAS 18264-1152 – techniques: spectroscopic}
   
   \titlerunning{NIR jets from a clustered region of massive star formation}
   \authorrunning{A. R. Costa Silva et al.}
   \maketitle
%

%
\section{Introduction} \label{intro}

The formation of massive stars ($M_*\gtrsim$\,8\,\msun) is a process that is not yet clearly understood. There are many observational challenges preventing us from uncovering the mechanisms behind their birth, such as the limited number of high-mass young stellar objects (HMYSOs), their location at large distances (typically a few kiloparsecs), and high visual extinction \citep[see, e.g.][for recent reviews]{tan2014, rosen2020}. We do know that, similarly to their low-mass counterparts, HMYSOs eject great amounts of material in the form of bipolar outflows and jets \citep[e.g.][]{frank2014,bally2016}. These outflows are intrinsically related to the process of accretion onto the protostellar surface \citep{blandford1982, pudritz1983,shu1994}. Thus, jets and outflows are crucial for pinpointing the location of massive protostars and providing insights into the physical processes unfolding in the central highly extinguished regions. 

When observing in the near infrared (NIR) regime, the driving source is usually not accessible as it is very often totally obscured. However, it is possible to observe the jets driven by massive young stars in the NIR and these can provide a wealth of information about the star-forming complex. Moreover, the excellent angular resolution usually achieved in this regime allows us to probe the collimated jet and individual knots. Molecular hydrogen (\hmol) is a particularly good shock tracer as it is the primary coolant in the NIR and its emission can be extended over spatial scales of several parsecs \citep{davis2004, caratti2008, davies2010, caratti2015, fedriani2018}, displaying numerous strong emission lines. In the \kband\ (1.9\,$-$\,2.5\,\micron) in particular, the strongest transition is the \hmol\ \mbox{1\,$-$\,0 S(1)} at 2.12 \micron, which has been used in many studies to probe the jet morphology in massive protostellar environments \citep{gredel2006, caratti2008}. HMYSO accretion is indeed revealed by their outflows, extending parsecs away from the star, but also through reflected light in their outflow cavity walls  \citep{fedriani2020}. This reflected light emission can further reveal the YSO nature when no clear association can be done with their outflows. In particular, the \brg\ at 2.16\,\micron\ is an excellent tracer of young protostars as it probes phenomena occurring very close to the YSO. The \brg\ line has been detected at the base of powerful jets driven by a HMYSO \citep{caratti2016}, as well as in accretion discs in a sample of HMYSOs \citep{koumpia2021}.

The target of our study is the \iras\ massive star-forming region. H$_2$O and Class I and II CH$_3$OH masers have been detected in this region \citep{sridharan2002, beuther2002d, chen2011, rodriguez2017}. Such masers are signposts of massive star formation and outflows. \iras\ is also associated with the extended green object (EGO) G19.88$-$0.53 \citep{cyganowski2008}, which has an estimated bolometric luminosity of 7.8\ee{3}\,\lsun\ from the Red Midcourse Space Experiment Source survey \citep[MSX,][]{lumsden2013}. The kinematic distance ambiguity to \iras\ was broken by \cite{roman2009}. \cite{he2012} locate it at 3.31$^{+0.34}_{-0.37}$\,kpc away and compute the local standard of rest velocity ($v_\mathrm{LSR}$) of the molecular cloud to be 43.7$\pm$0.02 \kms, which are the values adopted in our study.

In the past twenty years, this region has been surveyed in a variety of wavelengths, and there is growing evidence to support the scenario that \iras\ harbours more than one protostar. \cite{qiu2007} found two peaks in their 1.3\,mm and 3.4\,mm continuum maps. One of these peaks had been resolved as a triple source by \cite{zapata2006} in their Very Large Array (VLA) data at 1.3\,cm and 7\,mm. \cite{rosero2016} detected twelve compact radio sources in their sub-arcsecond resolution VLA observations at 6\,cm. In \cite{rosero2019}, the authors analysed eight of those sources (those within the EGO boundaries), suggesting that at least one could be an ionised thermal radio jet. This is in agreement with the analysis of \cite{zapata2006} for that source. The observations of \cite{issac2020} obtained with the upgraded Giant Meterwave Radio Telescope (uGMRT) at 1391.6\,MHz revealed two ionised thermal jets, one of which is associated with a hot dense core at 2.7\,mm as observed with the Atacama Large Millimeter/submillimeter Array (ALMA). In total, these authors found six 2.7\,mm cores in ALMA archival data consistent with being protostars with $M_*>9$\,\msun. The multi-wavelength analysis carried out by \citet{issac2020} strongly supports the protocluster hypothesis for this high-mass star-forming region.

Various outflows have also been observed in the \iras\ region via different molecular species and transitions. In the NIR regime, \citet{debuizer2010} detected EGO emission extending approximately 30\arcsec\ that was associated to \hmol\ outflow activity; \citet{varricatt2010} reported an east-west (EW) outflow extending at least 43.2\arcsec\, (i.e. $>0.69$\,pc at $d=3.31$\,kpc), with lobes apparently bent northwards (raising the possibility that they belong to different outflows), as well as additional \hmol\ features in the north-east (NE) and W directions; \citet{lee2012} found a 1.6$^\prime$\,(1.54\,pc) EW outflow, and additional knots to the NE and N of the region (which the authors speculate could be linked to multiple protostars); \citet{ioannidis2012} measured the E outflow lobe to be 42\arcsec\,(0.67\,pc), the W counterpart to be 50\arcsec\,(0.80\,pc), the N knot found in \citeauthor{ioannidis2012} to be 18\arcsec\,(0.29\,pc), and a new SW knot to be 10\arcsec\,(0.16\,pc); \citet{navarete2015} found a bipolar EW structure divided into six components, and three other non-aligned knots (one in the NE and two in the W directions). Summarising, there is wide agreement on the existence of two main \hmol\ lobes in the east and west directions. However, nothing is known about the kinematics of these flows and it remains unclear whether the \hmol\ knots compose one single outflow or if they belong to different outflows.

As for outflows in the sub-millimetre, millimetre, and radio regime, a CO (2$-$1) bipolar structure has been observed in the region of \iras\ by \citet{beuther2002b} with the IRAM 30\,m telescope at a resolution of 11\arcsec. A blue lobe was found to extend from E to W for $\sim$\,60\arcsec, with a SW red counterpart of $\sim$\,20\arcsec. The low angular resolution of these observations did not allow individual outflow structures to be resolved. More recently, \citet{issac2020} obtained low-resolution \mbox{C$^{18}$O (1$-$0)} data which revealed a large-scale NE\,-\,SW outflow consistent with the results of \citet{beuther2002b}. Moreover, this study presents high-resolution \mbox{C$^{18}$O (1$-$0)} and \mbox{C$^{17}$O (3$-$2)} observations, which traced a smaller scale collimated outflow in the SE\,-\,NW direction. The presence of outflows created by the shock tracer SiO (2$-$1) has been studied by \citet{qiu2007}, who found two quasi-perpendicular outflows in the SE\,-\,NW and in the NE direction \citep[with the former being consistent with the CO outflow reported in][]{issac2020}. These outflows showed several blue and red knots, leading to the proposition that there might be more than one protostar in the region. Indeed, \cite{issac2020} found six millimetric peaks, which hint at the presence of a protostellar cluster. \citet{sanchez2013} have also observed SiO (2$-$1), SiO (5$-$4), and HCO$^+$ (1$-$0) in the region of \iras\ using the IRAM 30m telescope (resolution from 10\arcsec\, to 30\arcsec), reporting bipolar outflow structures in all three cases.

In this paper, we present new VLT/SINFONI integral field unit observations of the central region with the highest angular resolution achieved so far (0.2\arcsec). Our goal is to study the morphology and investigate the line emitting regions of \iras\ in the NIR, with emphasis on the location and structure of the molecular hydrogen jets. We complement these observations with KMOS archival data in order to obtain a better understanding of the entire star-forming complex in the \mbox{\kband}. Furthermore, we present millimetre data of the CO (2$-$1) transition from the Submillimeter Array (SMA) in order to compare morphologies with the NIR \hmol\ outflows. Lastly, we create the spectral energy distribution (SED) for this source using archival data from \textit{Spitzer}, \textit{Herschel}, and WISE. The SED is fitted with radiative transfer models from \cite{zhangtan2018}, which are based on the Turbulent Core Accretion hypothesis for massive star formation \citep{mckeetan2003}. The best fit models provide constraints on the environmental conditions, evolutionary stage, and protostellar core mass for the protostar(s) in the \iras\ region.

This paper is structured as follows: in Sect. \ref{obs}, we present the observations and the data reduction processes; in Sect. \ref{results}, we describe the results found from the spectro-imaging analysis, the physical properties derived for the \hmol\ jets, the comparison to the CO outflows, and the SED fitting; we discuss the possible morphologies and jet structures in Sect. \ref{discussion}; finally, we summarise and highlight our conclusions in Sect. \ref{conclusions}.

\section{Observations and data reduction} \label{obs}

\begin{table*}
\caption{Summary of the observations for the region \iras.}
\label{tab:observations}      
\centering          
\begin{tabular}{c c c c c c c c}     
\hline\hline       
\noalign{\smallskip}
    Date of Obs. & Telescope & Wavelength/ & Total Exp. Time & FoV & Spectral Res. & Angular Res.\\ 
    (yyyy.mm.dd) & & Frequency & (s) & $\left( \arcsec \times \arcsec \right)$ & $(\mathrm{km\,s^{-1}})$ & $\left( \arcsec \right)$ \\
    \noalign{\smallskip}
    \hline              
    \noalign{\smallskip}
    2018-05-07 & VLT/SINFONI & $2.2$\,\micron & 900 & $8\times 8$ & 75 & 0.2\\
    \noalign{\smallskip}
    2013-09-28, &   &  &  &  &  &  \\  
    2013-07-01, & VLT/KMOS & $2.2$\,\micron & 4\,800 & $43.4\times 64.8$ & 71 & 0.6 \\  
    2013-06-29\, &   &  &  &  &  &  \\  
    \noalign{\smallskip}
    2006-07-11 & SMA & 1.3\,mm/230.5\,GHz & 15\,000 & $100\times100$ & - & 4.1$\times$\,3.2\\
\noalign{\smallskip}
\hline                  
\end{tabular}
\end{table*}

\subsection{VLT/SINFONI} \label{sinfoni}

We observed \iras\ with the Very Large Telescope (VLT) Spectrograph for Integral Field Observations in the Near Infrared \citep[SINFONI,][]{sinfoni1, sinfoni2} on 2018 May 7 (programme ID 0101.C-0317(A)), in the \kband\ (1.9 - 2.5 \micron), with a total exposure time of 900 seconds. The SINFONI set up yielded a field of view (FoV) of 8\arcsec$\times$8\arcsec, which was centred on coordinates \mbox{RA(J2000)\,=\,18:29:14.5}, \mbox{DEC(J2000)\,=\,-11:50:24.6}, with a position angle east of north of zero degrees. The spatial sampling was 0.125\arcsec$\times$0.250\arcsec pixel$^{-1}$, with the smaller sampling being along the northern direction. The adaptive optics (AO) system was utilised with the natural guide star 2MASS J18291556-1150164 (R$=$14.9\,mag), resulting in a spatial resolution of $\sim$\,0.2\arcsec\ and resolving power of R\,$\sim$\,4000 (i.e. spectral resolution of 0.55\,nm , corresponding to $\sim$\,75\,\kms).

The data were reduced with the EsoReflex SINFONI pipeline \citep[version 3.2.3,][]{esoreflex} and custom Python scripts. For telluric correction and flux calibration of the \iras\ observations, the telluric standard star HIP 88201 (spectral type B3V) was observed. Its spectrum was normalised with a blackbody function of the same temperature and a known \brg\ absorption feature was removed by fitting a Gaussian profile and subtracting it from the spectrum. The \iras\ data cube was wavelength calibrated by matching the peaks of several telluric lines to those in the ESO high-resolution \kband\ atmospheric spectrum\footnote{https://www.eso.org/sci/facilities/paranal/decommissioned/isaac/\\ tools/spectroscopic\_standards.html\#Telluric}. We also obtained the spectrum of a region of sky with no line emission from \iras\ and used it to subtract several OH lines associated with atmospheric features \citep{rousselot2000} which were not properly removed during the reduction process. A GAIA star (ID 4153125485659898624) lied in the SINFONI FoV and was used for accurate astrometric correction, so as to allow comparison with other data sets.

\subsection{VLT/KMOS} \label{kmos}

ESO archival data from the VLT \kband\ Multi Object Spectrograph \citep[KMOS,][]{kmos1, kmos2} in the \kband\ were retrieved for our target (science verification, programme ID 60.A-9458(A)). This instrument is formed of 24 integral field units (IFU), each with a FoV of 2.8\arcsec$\times$2.8\arcsec. When set into mosaic mode, the 24 arms can be combined in 16 pointings to create a map of $\sim$\,43.4\arcsec$\times$64.8\arcsec. The total exposure time of our KMOS observations was 4800 seconds (16 pointings times 300\,s exposure), the spatial sampling is 0.2\arcsec$\times$0.2\arcsec, with a seeing-limited angular resolution of 0.6\arcsec, and resolving power in the \kband\ of R\,$\sim$\,4200 (i.e. spectral resolution $\sim$\,0.52\,nm, corresponding to $\sim$\,70\,\kms). 

The data are composed of three data sets, each spanning a different part of the complex. The wider outflow area was observed on 2013 June 29 and 2013 July 1, and the central region was observed on 2013 September 28. The plots presented throughout the paper are combinations of the three observations. However, since these data sets belonged to a science verification programme, some issues were encountered, such as the central region not being fully covered. The missing regions are represented as white or black boxes in the figures showing KMOS data. In fact, when KMOS and SINFONI observations overlap, the latter should take precedence in the results. 

The data were reduced using the EsoReflex KMOS pipeline (version 3.0.1) and custom Python scripts. The data were flux calibrated using stars in the field of view that are also present in the 2MASS and UKIDSS catalogues. The uncertainty in flux calibration is estimated to be between 5\% and 10\%, and it results from combining the photometric errors of the catalogues and those from the data. The stars showed little variation in the measured magnitudes. We also extracted a region of the sky with no emission lines to remove any remaining OH residuals from the data reduction process.

\subsection{SMA} \label{sma}

The Submillimeter Array (SMA) was used on June 17 and August 09, 2006 to map the $J=2\rightarrow 1$ CO rotational transition ($\sim$230.5 GHz) towards \iras. Observations were performed using the SMA extended and compact configurations of the interferometer. For the observations in the compact configuration, only seven antennas were available. In June (extended configuration), the observations were performed under good weather conditions with zenith opacities $\tau$(225\,GHz) between 0.1 and 0.14 measured by the National Radio Astronomy Observatory (NRAO) tipping radiometer operated by the Caltech Submillimeter Observatory (CSO). In August, the weather conditions were less favourable, with $\tau$(225\,GHz) between 0.15 and 0.2.

Bandpass calibration was done with 3C454.3 and 3C273. We used Callisto and Uranus for the flux calibration which is estimated to be accurate within 20\%. Gain calibration was done via frequent observations of the quasars 1911-201 and 1833-21. The initial flagging and calibration was done with the IDL superset MIR originally developed for the Owens Valley Radio Observatory and adapted for the SMA. Images were produced using uniform weighting. The final map has phase centre in \mbox{RA(J2000)\,=\,18:29:14.55}, \mbox{DEC(J2000)\,=\,-11:50:22.50}, and a synthesised beam of 4.1\arcsec$\times$\,3.2\arcsec. Table\,\ref{tab:observations} summarises the observations presented in this paper.

\section{Analysis and results} 
\label{results}

We take advantage of the IFU nature of the SINFONI and KMOS instruments to thoroughly examine the \iras\ region through spectro-imaging analysis. As stated before, our main focus are the NIR \hmol\ jets, as their geometry provides clues to the location of the driving sources. The \hmol\ emission lines allow us to study the radial velocities of the jet knots and the excitation conditions, as well as obtain estimates for visual extinction, \hmol\ temperatures, \hmol\ column densities, and \hmol\ knot masses. The SMA CO data is used to peer into the regions where the NIR is obscured and to confirm outflow emission and kinematics. We detail our analysis and present our results in this section, and further discuss their implications in Sect. \ref{discussion}.

\subsection{Point sources in SINFONI and KMOS FoV}
\label{spectral_analysis}

\begin{figure*}
    \centering
    \includegraphics[width=0.48\hsize]{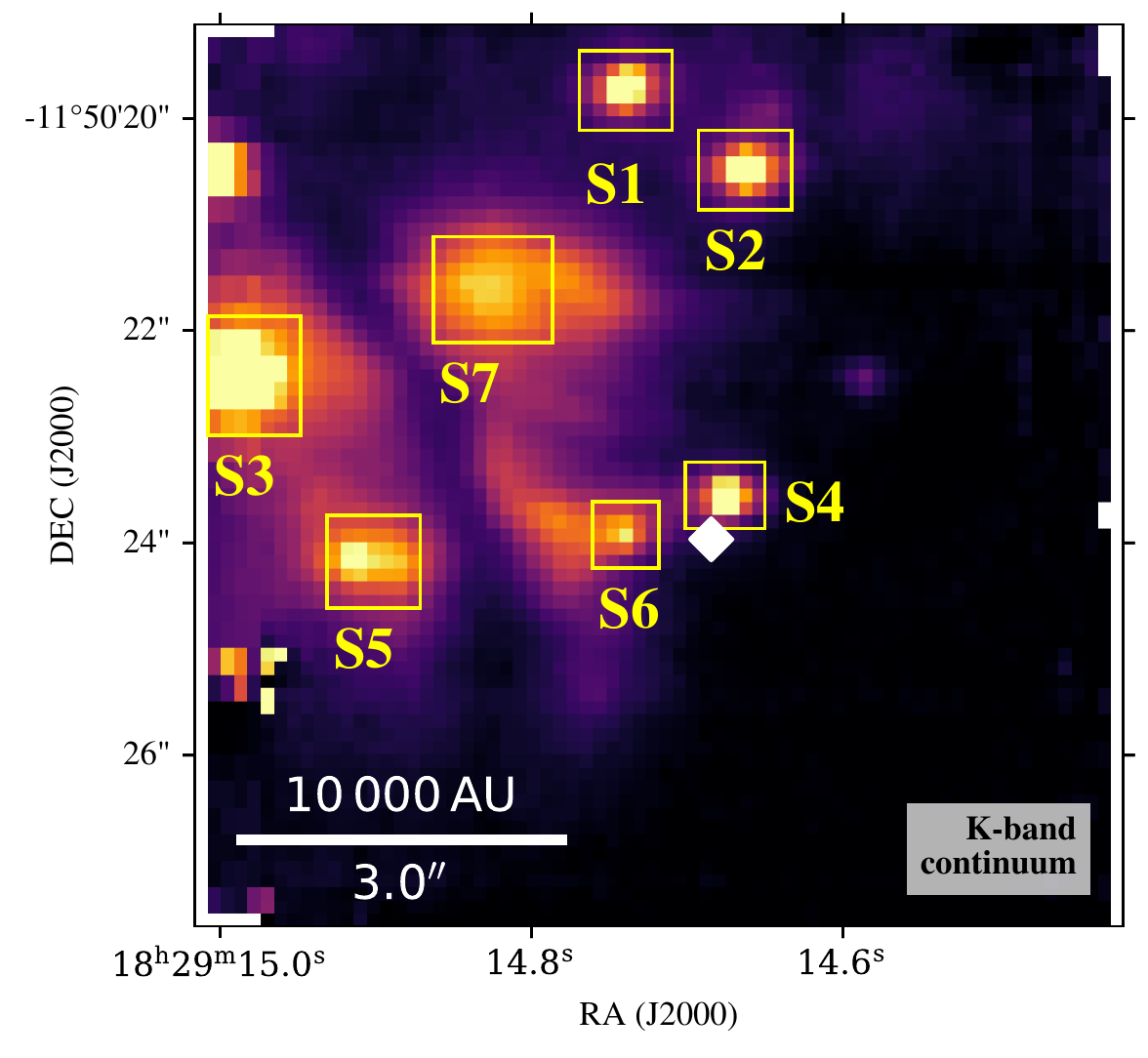}
    \includegraphics[width=0.48\hsize]{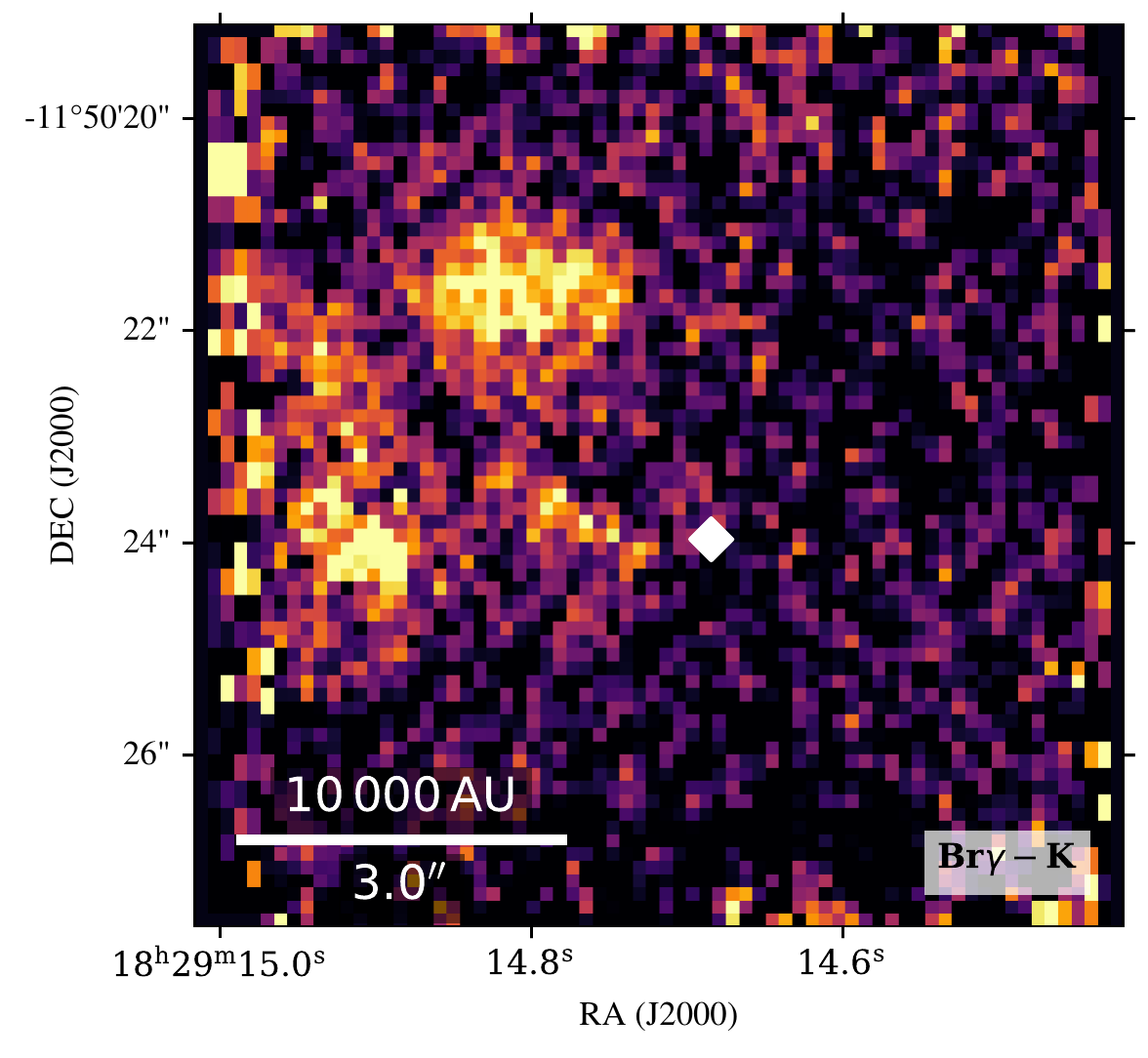}
    \includegraphics[width=0.94\hsize]{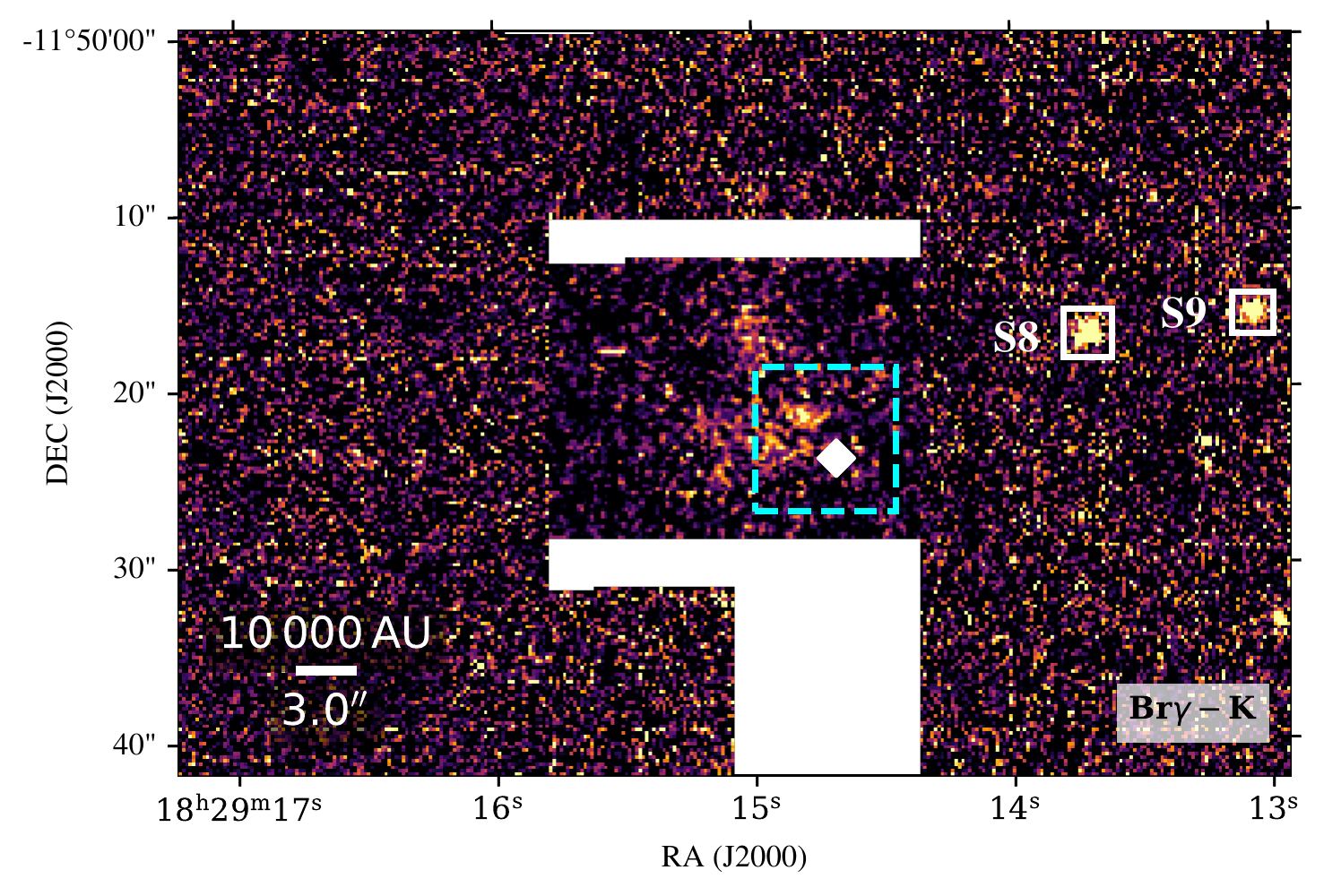}
    \caption{ \textit{Top left:} SINFONI \kband\ continuum emission. We identify seven point sources for which spectra are shown in Fig. \ref{fig:point_sources_spectra}. \textit{Top right:} Continuum-subtracted \brg\ emission map in the SINFONI FoV. \textit{Bottom:} Continuum-subtracted \brg\ emission map in the KMOS FoV, where two bright sources are identified. The white regions were not covered by the observations. The cyan dashed square represents the SINFONI FoV. The white diamond indicates the position of the \iras\ source (RA(J2000)=18:29:14.6846, Dec(J2000)=$-$11:50:23.966, taken from SIMBAD).}
    \label{fig:continuum_image}
\end{figure*}

\begin{figure*}
    \centering
    \includegraphics[width=0.92\hsize]{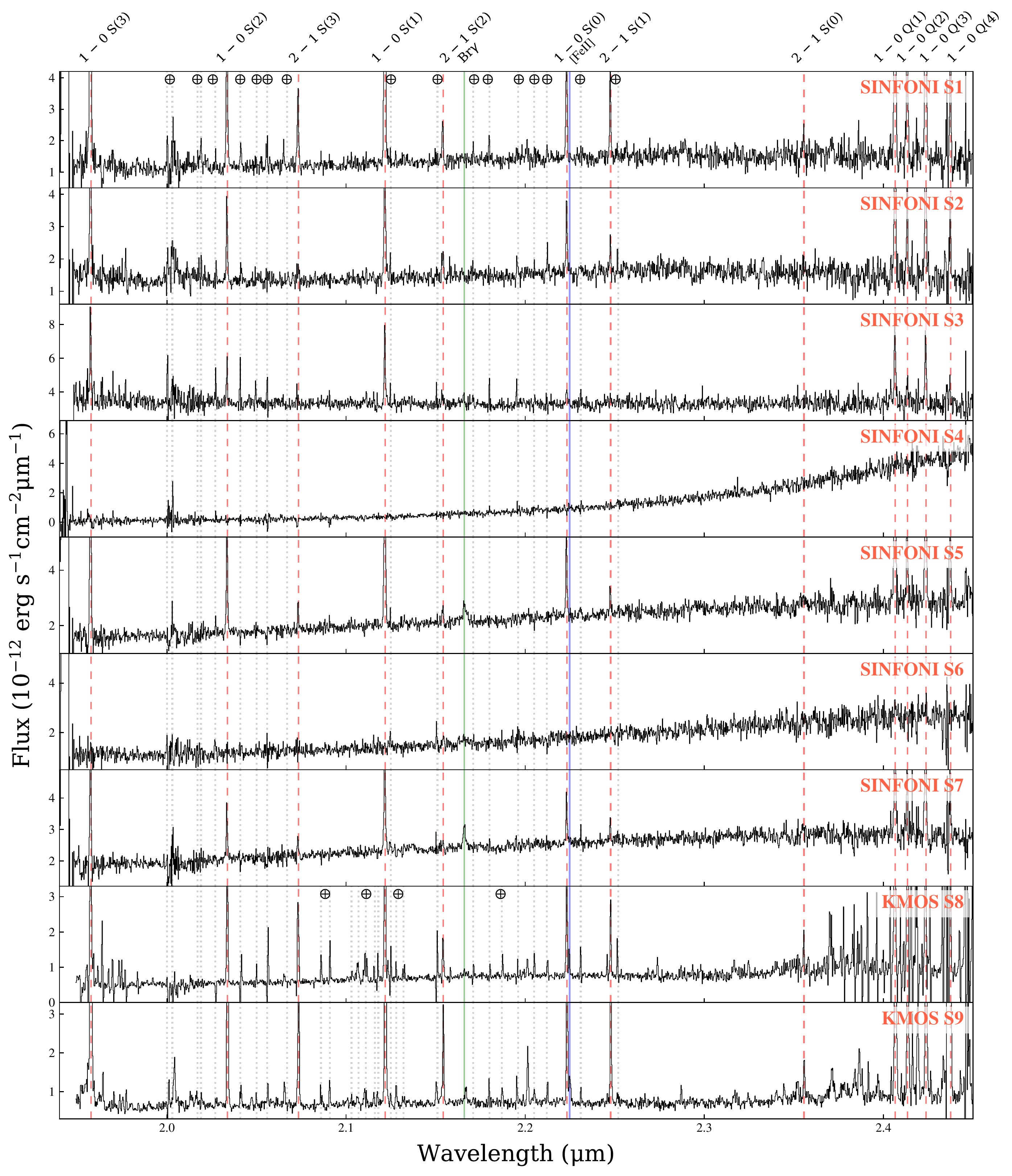}
    \caption{\kband\ spectra of point sources in SINFONI (S1\,-\,S7) and KMOS (S8\,-\,S9). Strong lines are cut off by the figure limits to allow for the identification of weaker features. The dashed red lines reveal the \hmol\ transitions, the full green line identifies the \brg\ transition, the full blue line identifies the \feii transition. The Earth symbols and dotted grey lines label the residual atmospheric features (i.e. OH lines).}
    \label{fig:point_sources_spectra}
\end{figure*}

We present the SINFONI \kband\ continuum emission\footnote{The KMOS \kband\ continuum image contains many artefacts due to the data belonging to a science verification programme. It is shown in Fig. \ref{fig:kmos_continuum}.} in the upper left panel of Fig. \ref{fig:continuum_image}, created using 359 line-free continuum channels from across the entire \kband. We identified seven point sources in this image (designated from S1 to S7) for which \kband\ magnitudes were derived and are given in Table \ref{tab:magnitudes}. Source S3 coincides with the GAIA detection used for accurate astrometric correction (see Sect.\,\ref{sinfoni}), enabling comparison with radio data from the literature (see Sect. \ref{discussion}). The spectra extracted at the location of the point sources are shown in Fig. \ref{fig:point_sources_spectra}. In the spectra, we observed several \hmol\ lines (from the 1\,$-$\,0 S(3) at 1.95 \micron\ to the 1\,$-$\,0 Q(4) at 2.43 \micron, see Sect. \ref{h2_emission} and Table \ref{tab:h2_lines}), as well as the \ion{H}{I} recombination line \brg\ (2.166 \micron), and the forbidden iron transition \feii\ ($^2H_{11/2}-^2G_{9/2}$, 2.225 \micron) associated with shocked emission. The spectra also show some OH telluric residuals. All of these features are labelled in Fig. \ref{fig:point_sources_spectra}.

\begin{table}
    \caption{\kband\ magnitudes and \brg\ fluxes of point sources.}
    \label{tab:magnitudes}
    \centering
    \begin{tabular}{lcc}
        \hline \hline
        \noalign{\smallskip}
         Source & Magnitude & \brg\, flux \\
         & (mag) & (10$^{-16}$ erg s$^{-1}$ cm$^{-2}$) \\  
        \noalign{\smallskip}
        \hline
        \noalign{\smallskip}
         S1 & 13.7\,$\pm$\,0.1 & - \\
         S2 & 13.6\,$\pm$\,0.2 & - \\
         S3 & 12.8\,$\pm$\,0.1 & - \\
         S4 & 14.8\,$\pm$\,0.3 & - \\
         S5 & 13.3\,$\pm$\,0.1 & 11.8\,$\pm$\,0.9 \\
         S6 & 13.6\,$\pm$\,0.1 &  6.9\,$\pm$\,1.3 \\
         S7 & 13.1\,$\pm$\,0.1 & 12.4\,$\pm$\,1.2 \\
         S8 & 14.6\,$\pm$\,0.2 &  2.6\,$\pm$\,0.6 \\
         S9 & 14.5\,$\pm$\,0.1 &  4.9\,$\pm$\,0.1 \\
        \noalign{\smallskip}
        \hline
    \end{tabular}
    \tablefoot{Sources identified in the SINFONI (S1$-$S7) and KMOS (S8$-$S9) field of view (see Fig. \ref{fig:continuum_image}).}
\end{table}

We further investigate the emission of the accretion and ejection tracer \brg\ by analysing the continuum-subtracted emission in the SINFONI and KMOS fields of view (the corresponding emission maps are shown in the upper right and bottom panel of Fig. \ref{fig:continuum_image}). In the SINFONI map, \brg\ emission is found in the locations of the point sources S5, S6, and S7, which also show a rising continuum. The slightly extended emission resembles that of the continuum map, meaning it is most likely scattered light being reflected on the outflow cavities of young sources. An additional source with a rising \kband\ continuum, S4, is identified in our spectra (see Fig. \ref{fig:point_sources_spectra}). In contrast with the others, S4 does not show any \brg\ or \hmol\ lines. These characteristics suggest that the source is very embedded and might be a young source. In the KMOS map, we identify two point sources (labelled S8 and S9) which have strong \brg\ emission. These young sources are also present in the KMOS \kband\ continuum map (Fig. \ref{fig:kmos_continuum}). They might belong to the \iras\ cluster, or they could be unrelated sources which happen to lie along the line of sight (see Sect. \ref{discussion}). The spectra for S8 and S9 are included in Fig. \ref{fig:point_sources_spectra} alongside the SINFONI point sources spectra. We suggest that S4, S5, S6, S7, S8, and S9 may be YSOs revealed by our NIR observations. We give the line fluxes of \brg\ for all emitting sources in Table \ref{tab:magnitudes}.
%

\begin{figure*}
    \centering
    \includegraphics[width=0.94\hsize]{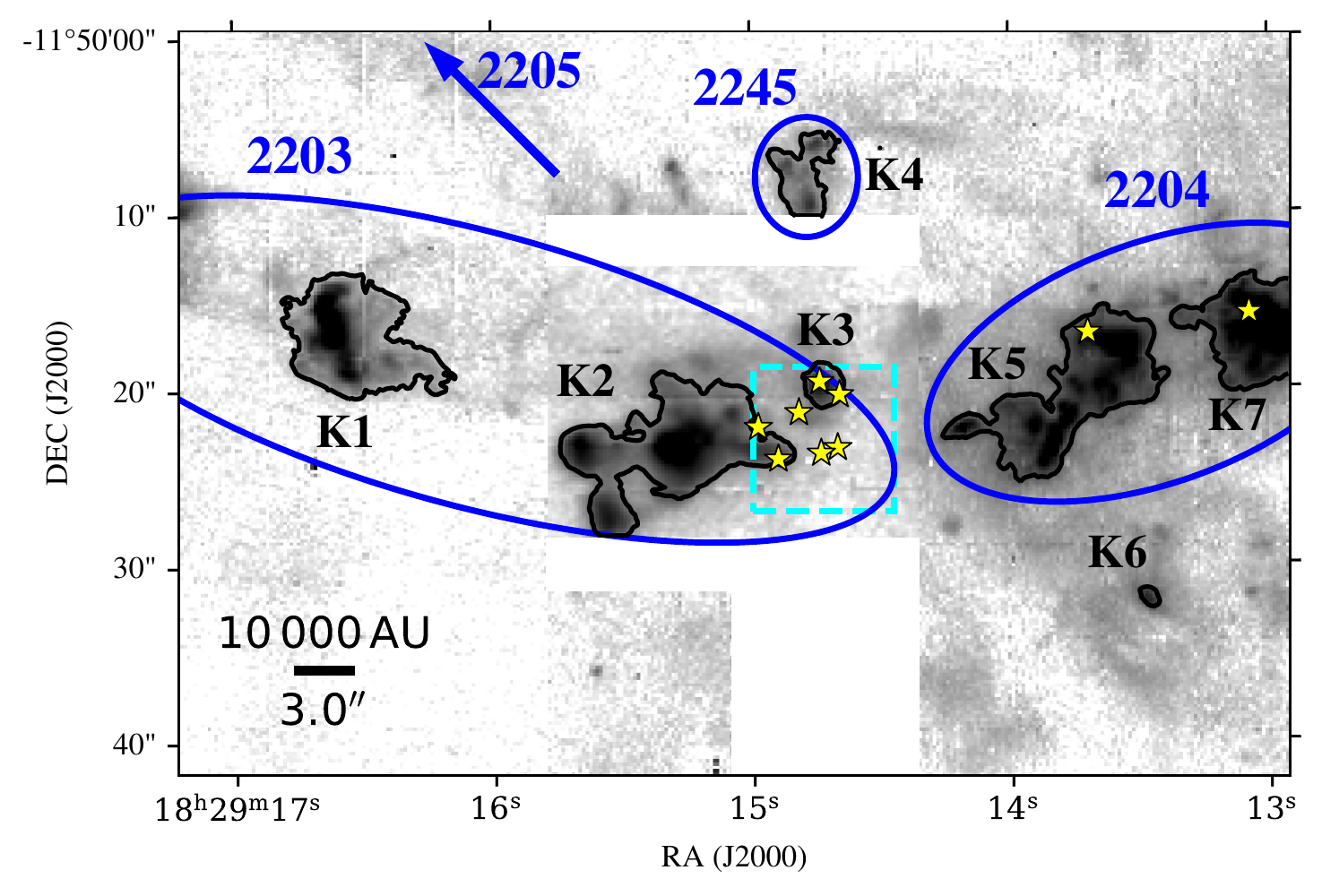}
    \includegraphics[width=0.5\hsize]{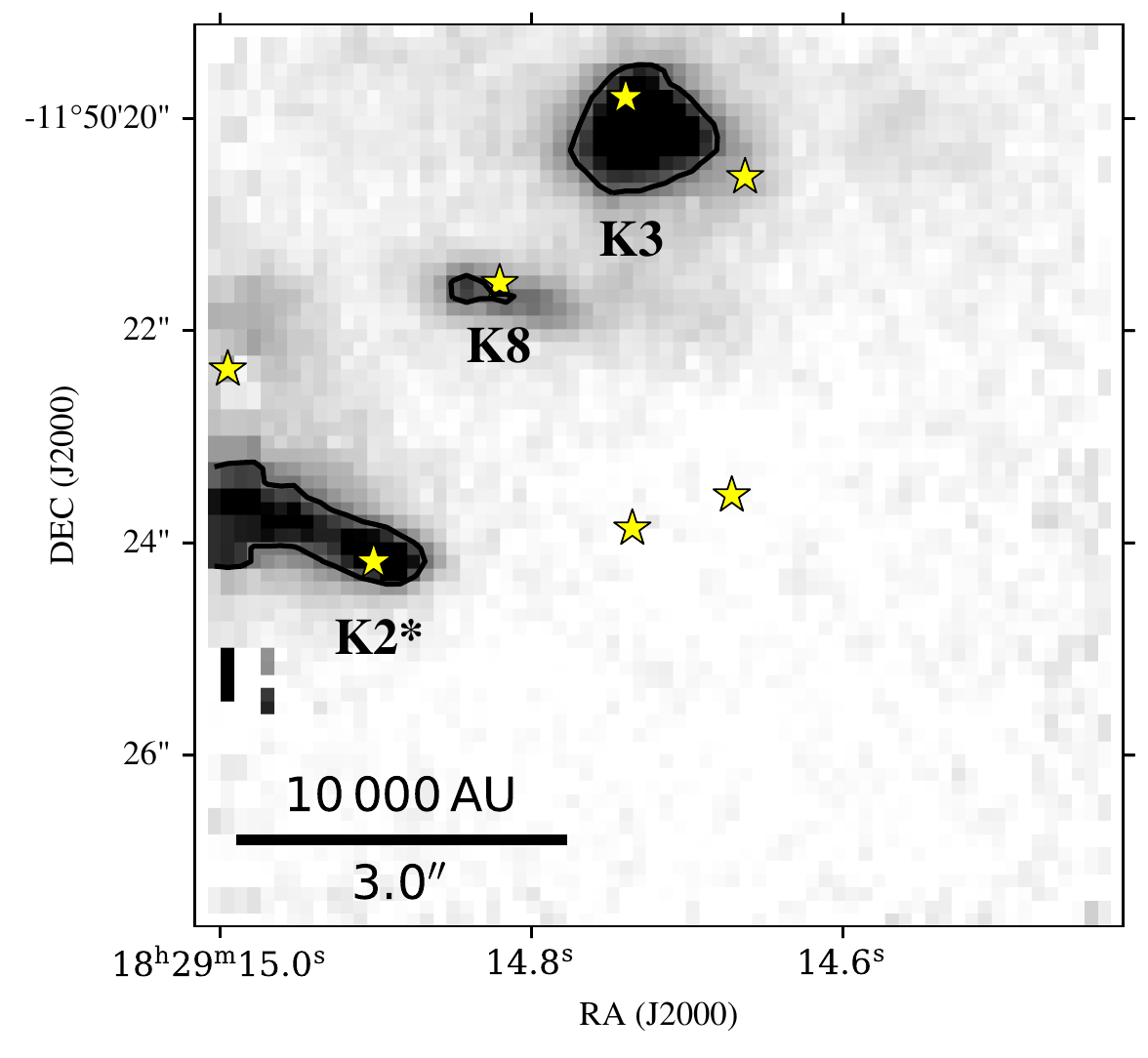}
    \caption{\hmol\ 1\,$-$\,0 S(1) (2.12 \micron) continuum-subtracted emission, with jet knots outlined according to a 3$\sigma$ threshold of the local background. The yellow stars represent the point sources reported in Sect. \ref{spectral_analysis}. \textit{Top panel:} KMOS observations. The ellipses correspond to the molecular hydrogen objects (MHOs) in the \iras\ region; the white rectangles were not covered by the observations; the dashed cyan rectangle defines the SINFONI field of view. \textit{Bottom panel:} SINFONI observations.}
    \label{fig:h2_image}
\end{figure*}

\subsection{\hmol\ emission}
\label{h2_emission}

Figure \ref{fig:h2_image} shows the continuum-subtracted \hmol\ emission from the \mbox{1\,$-$\,0 S(1)} transition, at 2.12 \micron, observed by KMOS (top panel) and SINFONI (bottom panel). The wider field of view provided by KMOS reveals a large \hmol\ outflow in the east-west direction, with several jet knots and bow shocks, as well as some emission to the north of the region. These knots correspond to the Molecular Hydrogen Objects (MHOs) 2203, 2204, and 2245 first identified in \cite{varricatt2010} and in \cite{lee2012}. Our KMOS emission map also reveals some further extended emission towards the south-west direction and towards the north-east region (a possible link to MHO 2205), although this emission is slightly below our 3$\sigma$ detection limit.

Though the east and west lobes are cut-off due to our field of view, we estimate these outflows extend for at least $\sim$\,41\arcsec (i.e. $>0.67$\,pc at $d=3.31$\,kpc) and $\sim$\,22\arcsec ($>0.35$\,pc) with opening angles of $\sim$\,11$^\circ{}$ and $\sim$\,22$^\circ{}$, respectively. The northern emission extends for $\sim$\,17\arcsec ($>0.27$\,pc), and strong \hmol\ emission to the south-west is found at a distance of $\sim$\,17\arcsec ($>0.27$\,pc) from the central region.

We define the limits of the knots in both data sets using a 3$\sigma$ threshold of the local background. This method outlines seven knots (labelled from K1 to K7) in the KMOS data, and three jet knots (K2*, K3, K8) in the SINFONI image. Knot K2* is part of the K2 knot seen by KMOS, K3 is also seen by KMOS, whereas K8 is barely visible in the KMOS data and is only resolved by SINFONI due its higher angular resolution.

Using the \hmol\ emission from these knots, we have made several maps to investigate the physical properties of the \hmol\ emitting region and its origin. In the next subsections we present: the velocity, excitation, and extinctions maps, as well as the ro-vibrational diagrams and properties derived for the various knots.

\subsubsection{Velocity maps} 
\label{velocity_maps}

\begin{figure*}
    \centering
    \includegraphics[width=0.94\hsize]{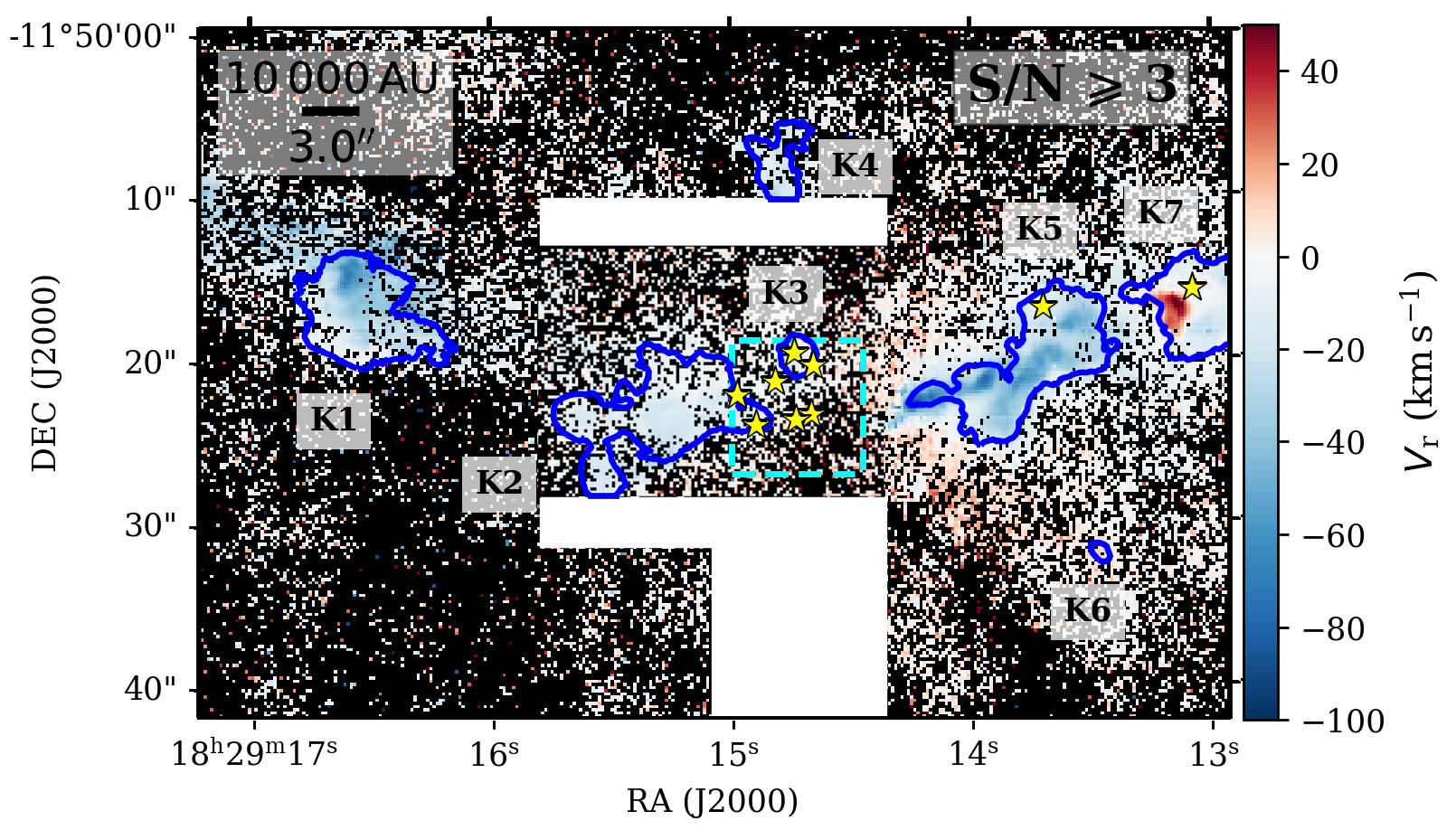}
    \includegraphics[width=0.5\hsize]{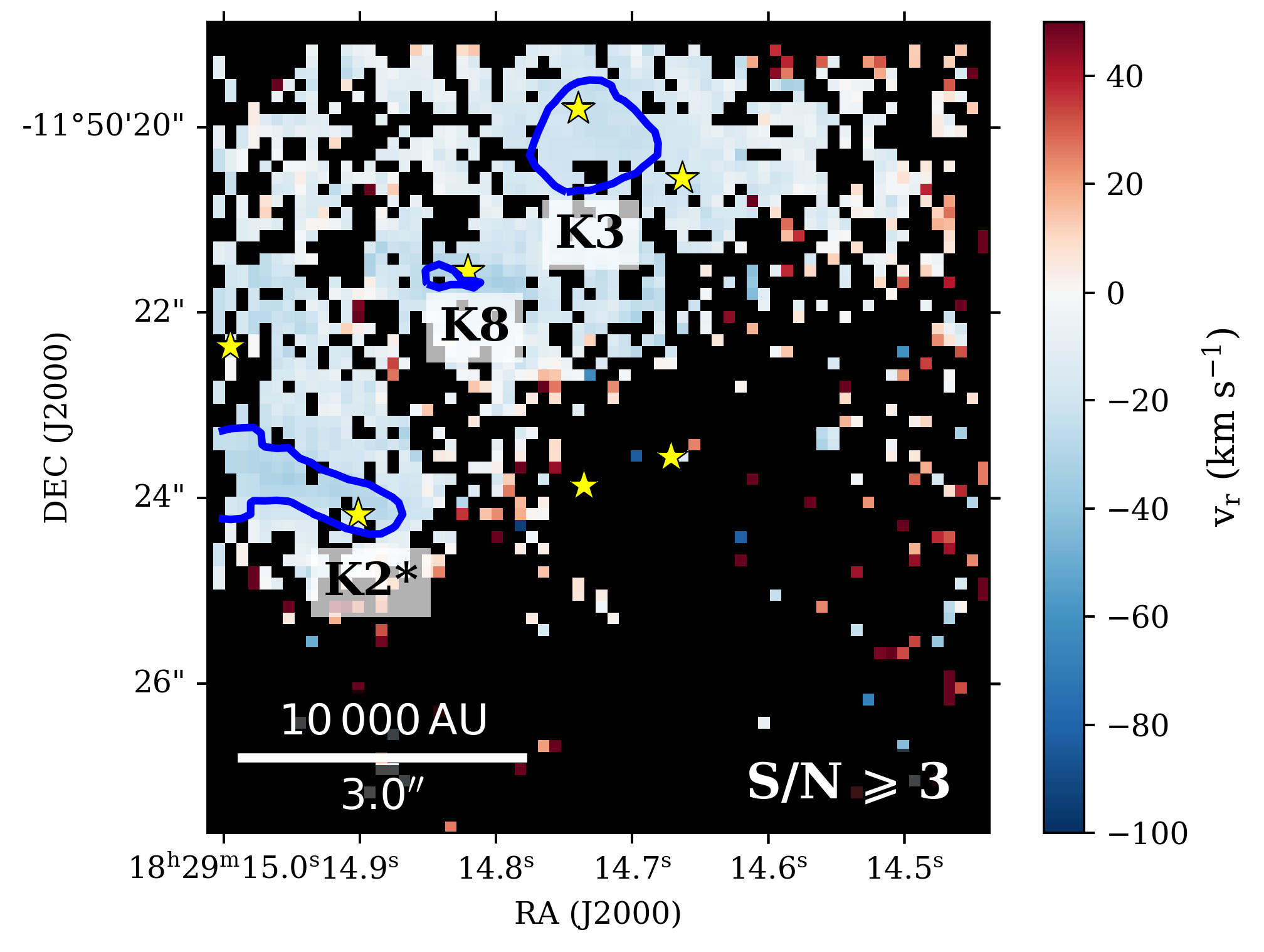}
    \caption{Radial velocity ($v_r$) maps obtained via measuring the Doppler shift in the \hmol\ 1\,$-$\,0 S(1) (2.12 \micron) line. The velocities are in \kms\ and in the local standard rest velocity and corrected by the velocity of the cloud \citep[$v^\mathrm{cloud}_\mathrm{LSR} = 43.7 \pm 0.02$ \kms,][]{he2012}. The yellow stars represent the point sources reported in Sect. \ref{spectral_analysis}, and the contours are the \hmol\ knots (same as Fig. \ref{fig:h2_image}). \textit{Top panel:} KMOS observations. The white regions were not covered by the observations. The dashed cyan rectangle defines the SINFONI field of view. \textit{Bottom panel:} SINFONI observations.}
    \label{fig:radial_velocity}
\end{figure*}

We produced radial velocity ($v_r$) maps from the KMOS and SINFONI data by fitting the \hmol\ \mbox{1\,$-$\,0 S(1)} (2.12 \micron) line with a Gaussian profile and measuring the Doppler shift of the line peak. The values are given with respect to the local standard of rest (LSR) and corrected by the velocity of the cloud \citep[$v^\mathrm{cloud}_\mathrm{LSR}\,=\,43.7\,\pm\, 0.02$\,\kms,][]{he2012}. The maps are shown in Figure \ref{fig:radial_velocity}. Even though the nominal resolution of our observation is $\sim70$\,\kms, we can measure more accurately the $v_r$ for the high signal-to-noise ratio (S/N) lines. Therefore, the precision in velocity of our measurements (typically calculated as $\mathrm{Resolution}/\sqrt{\mathrm{S/N}})$ is $\sim$\,20\,\kms, taking an average S/N$\sim$\,10.

The map reveals that the two lobes of the large-scale E-W outflow seen in Fig. \ref{fig:h2_image} are both blue-shifted, with radial velocities ranging from $\sim\,-$20\,\kms\ to $\sim\,-$80\,\kms. Under the assumption that outflows are symmetrically bipolar, this indicates that MHO 2203 (i.e. knots K1, K2, K8, maybe K3) and MHO 2204 (i.e. knots K5, K7) actually belong to different outflows, but the red-shifted counterparts of these lobes are not as evident in the radial velocity map nor in the intensity map. They might be obscured and the signal-to-noise is too low to allow for a meaningful detection. However, we note that the south-west region of KMOS is tentatively red-shifted, with the average radial velocity measured within the limits of K6 being \mbox{$\sim$\,10\,\kms}. Thus K6 could be a counterpart for MHO 2203.

Given that we have lower limits on the lengths of MHOs 2203 and 2204, we can take mean values of radial velocities and estimate values of outflow ages for certain inclination angles ($i$, defined as the angle between the plane of the sky and the outflow axis). For MHO 2203, we obtain an age of 74.3\,kyr assuming $i=30^{\circ}$, and 222.9\,kyr if $i=60^{\circ}$. In the case of MHO 2204, the age estimate is 28.5\,kyr for $i=30^{\circ}$, and 85.4\,kyr for $i=60^{\circ}$.

The northern knot K4 is blue-shifted ($\sim\,-$15\,\kms) and its alignment with knot K3 suggests that both of them might belong to a third outflow being driven northwards. To the west of the KMOS field of view, a clearly red-shifted bow shock ($\sim$\,40\,\kms) is revealed within K7, indicating that this region likely harbours a separate driving source of a bipolar outflow. The geometry of the bow shock indicates that its driving source is towards the SE direction, which could be S8 or another source from the central region of the cluster.

\subsubsection{Excitation maps} \label{excitation_maps}

\begin{figure*}
    \centering
    \includegraphics[width=0.94\hsize]{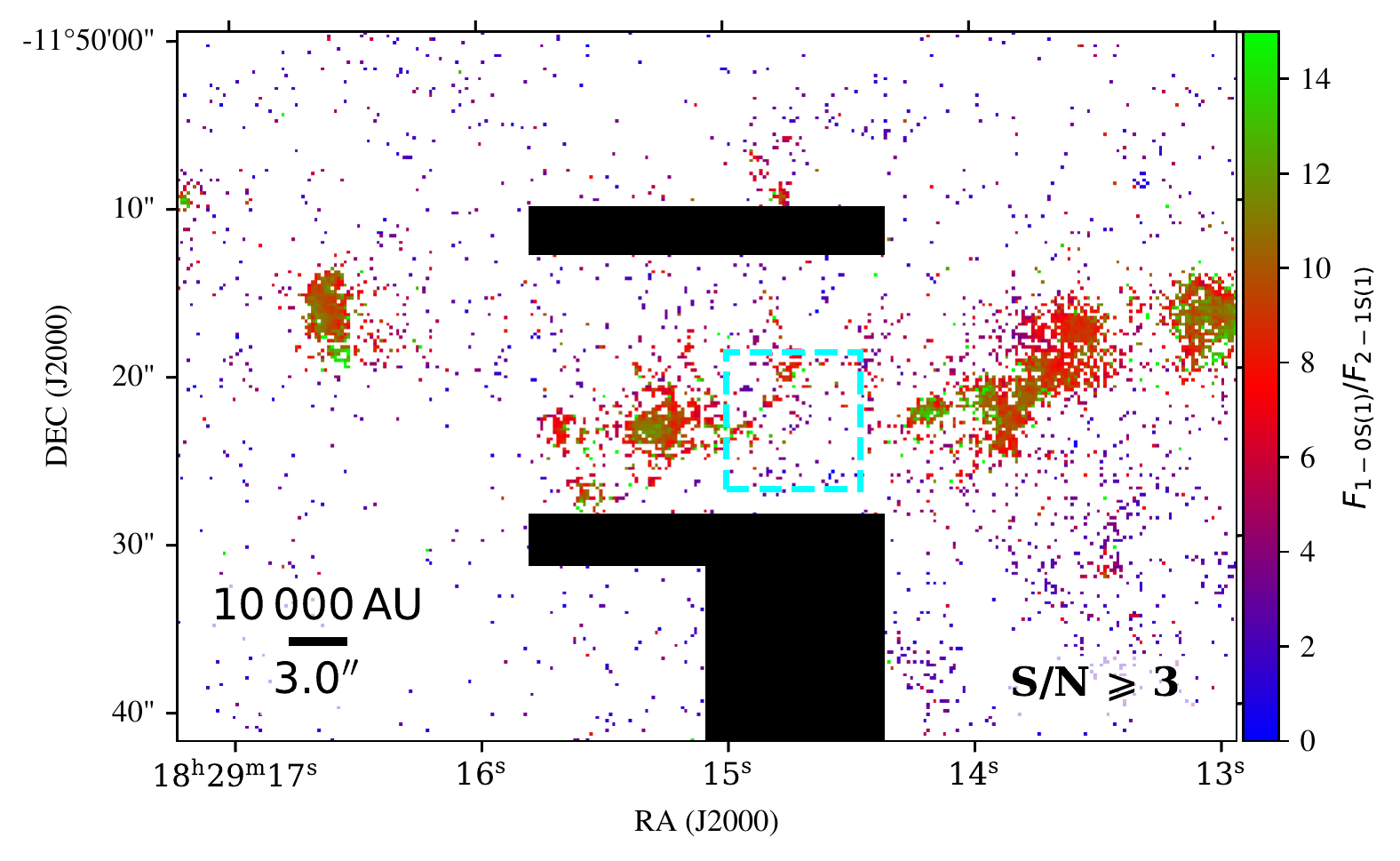}
    \includegraphics[width=0.5\hsize]{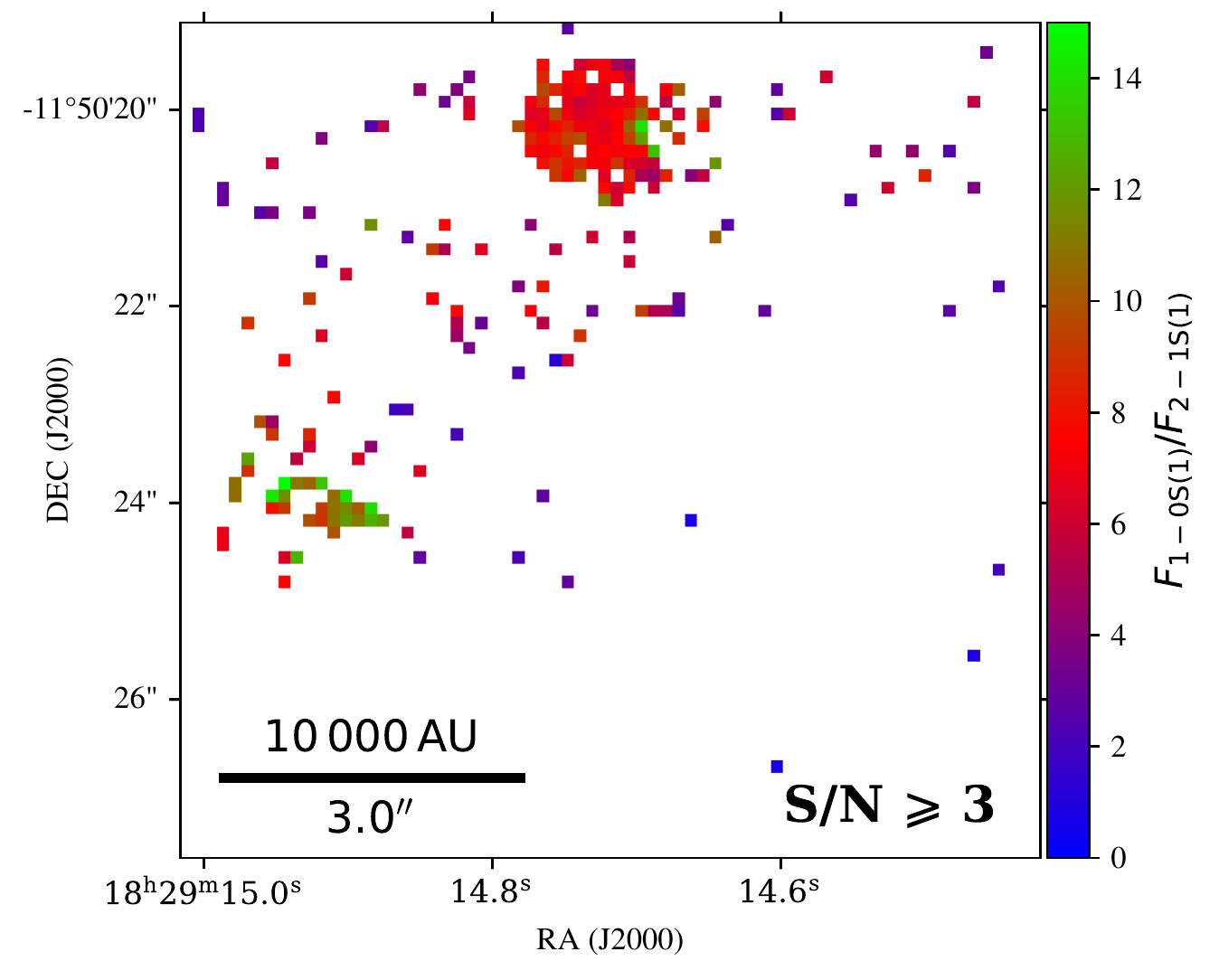}
    \caption{Excitation maps. \textit{Top panel:} KMOS observations. The black regions were not covered by the observations. The dashed cyan rectangle defines the SINFONI field of view. \textit{Bottom panel:} SINFONI observations.}
    \label{fig:excitation_map}
\end{figure*}

We investigate the nature of the \hmol\ emission to inspect whether it is shocked or fluorescent emission. The excitation maps presented in Fig. \ref{fig:excitation_map} are obtained by calculating the integrated line flux ratio of the \hmol\ transitions \mbox{1\,$-$\,0 S(1)} and \mbox{2\,$-$\,1 S(1)} at 2.12 \micron\ and 2.25 \micron, respectively (i.e. $F_\mathrm{1-0 S(1)}/F_\mathrm{2-1 S(1)}$). Only the pixels where both lines were fitted with a Gaussian profile with a $S/N\,>$\,3 are considered for this analysis. Typically, \mbox{1\,$-$\,0 S(1)}/\mbox{2\,$-$\,1 S(1)} ratios larger than 3 indicate shocked emission, whereas values $\leq 3$ indicate fluorescent emission \citep[see, e.g.][]{burton1992, wolfchase2017}.

All of the knots show ratios between 8 and 12 which are values consistent with shocked emission \citep[see also][]{oh2016, wolfchase2017}. We note that this ratio by itself is not sufficient to categorise the emission as definitely shock driven. However, given the morphologies and radial velocities of the \hmol\ emission features, the correspondence with other outflow tracers, and the thermal distribution seen in the ro-vibrational diagrams (see below), it is most likely that we are observing \hmol\ shocked material.

Notably, none of the \hmol\ emitting regions in the SINFONI or KMOS maps show any indication of fluorescent emission. Even the extended emission around S5, S6, and S7 in SINFONI data does not show any such indication. This supports the idea that these sources may be low-mass YSOs. Moreover, the lack of \hmol\ fluorescent emission in the region indicates that the massive young stars present there are either too embedded to excite the \hmol, too young, or both.

\subsubsection{Extinction estimates and physical properties of jets}

\begin{figure*}
    \centering
    \includegraphics[width=0.94\hsize]{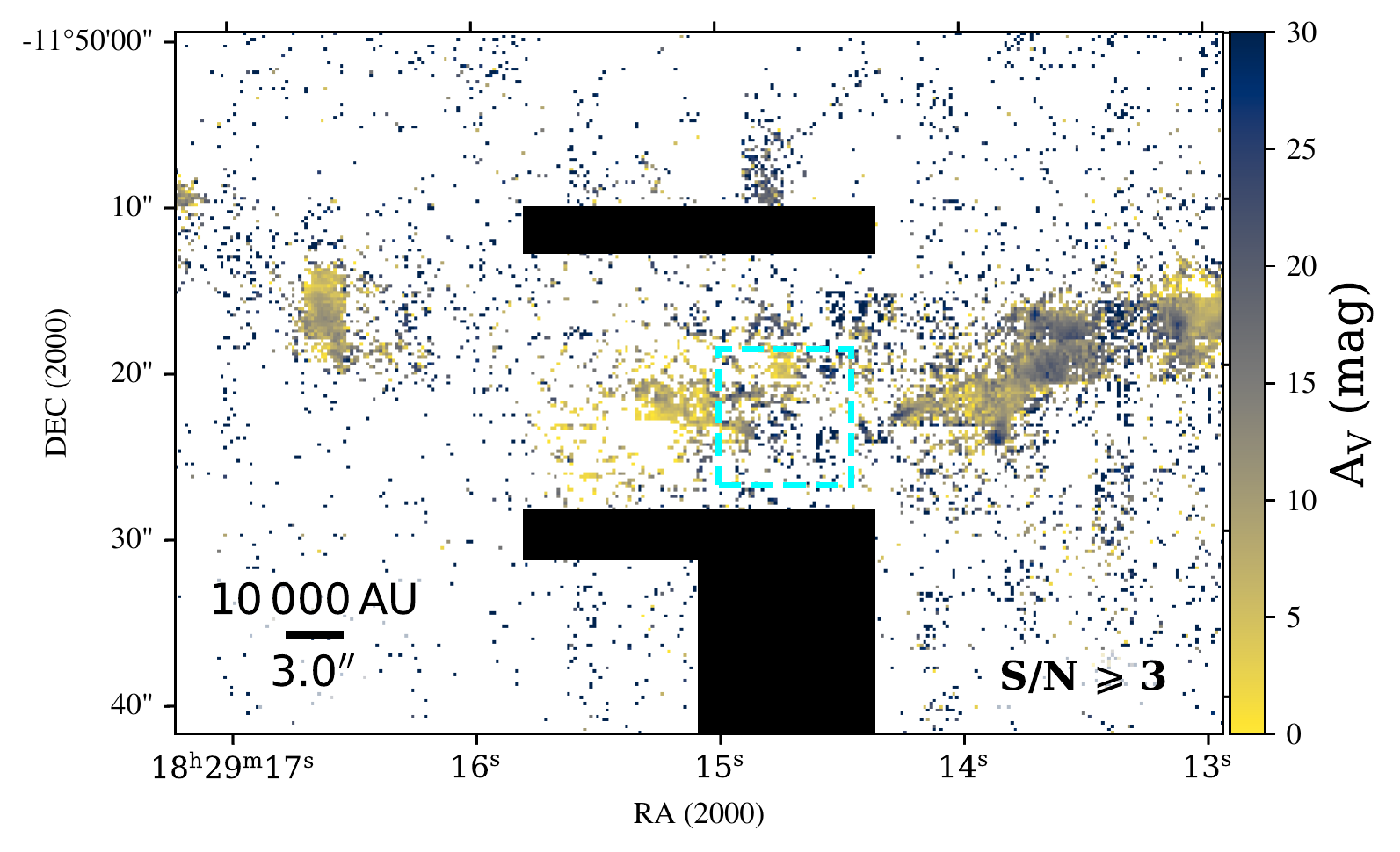}
    \includegraphics[width=0.5\hsize]{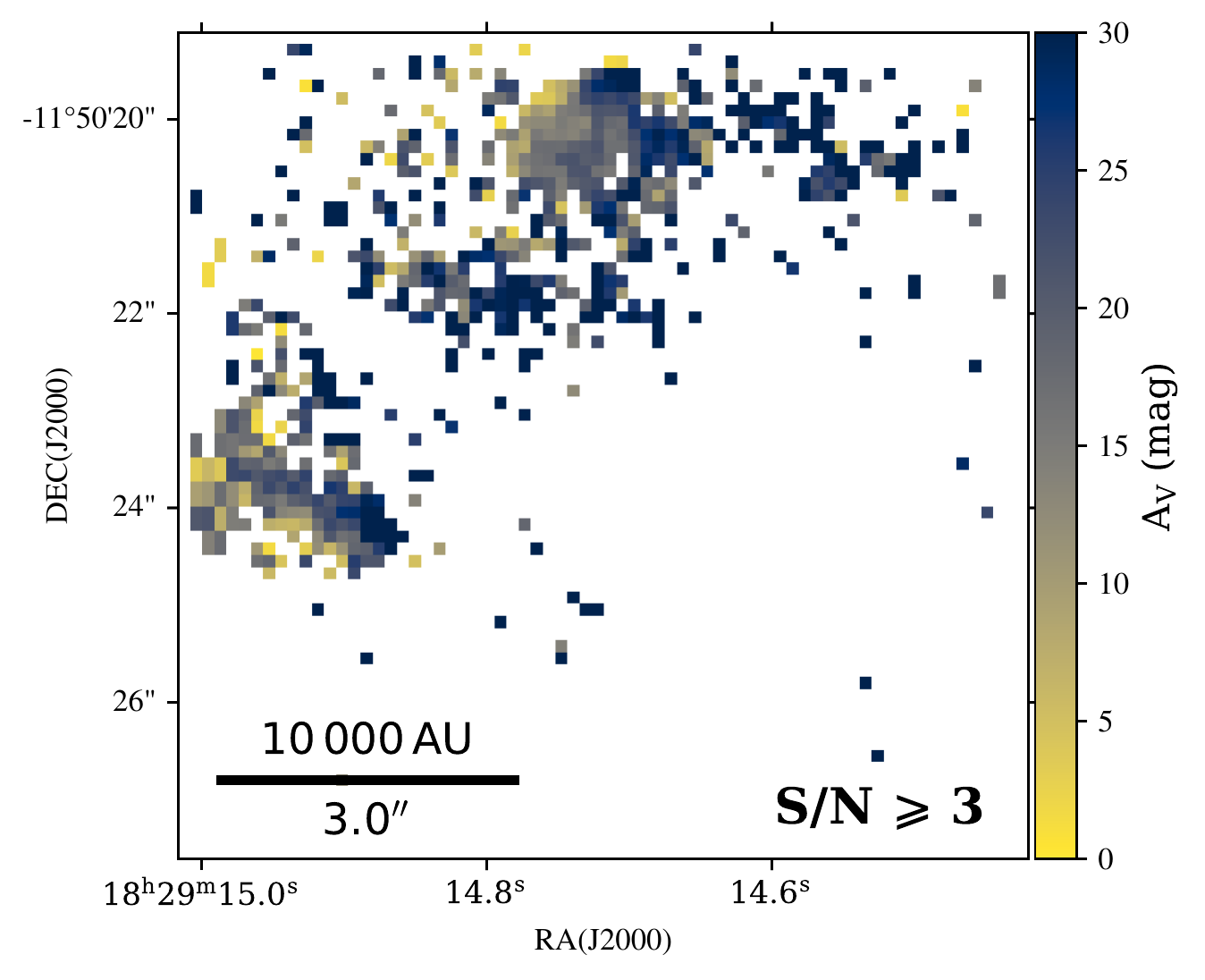}
    \caption{Excitation maps using the ratio of the \hmol\ lines \mbox{1\,$-$\,0 S(1)} and \mbox{2\,$-$\,1 S(1)} at 2.12 \micron\ and 2.25 \micron. \textit{Top panel:} KMOS observations. The black regions were not covered by the observations. The dashed cyan rectangle defines the SINFONI field of view. \textit{Bottom panel:} SINFONI observations.}
    \label{fig:extinction_map}
\end{figure*}

We calculated visual extinction (\av) estimates for the \iras\ region using the integrated line flux ratios of the \hmol\ lines \mbox{1\,$-$\,0 S(1)} and \mbox{1\,$-$\,0 Q(3)} (i.e. $F_{\lambda_\mathrm{1}}$, $F_{\lambda_\mathrm{2}}$). For details of the method used, we refer the reader to Appendix \ref{extinction_appendix}. The visual extinction maps can be seen in Fig. \ref{fig:extinction_map}. The knots mostly show \av\ between $\sim$\,10 and $\sim$\,20 magnitudes (see Table 3). With these \av\ estimates, we are then able to evaluate other important properties of the jets, as explained below.

\begin{figure}
    \centering
    \includegraphics[width=\hsize]{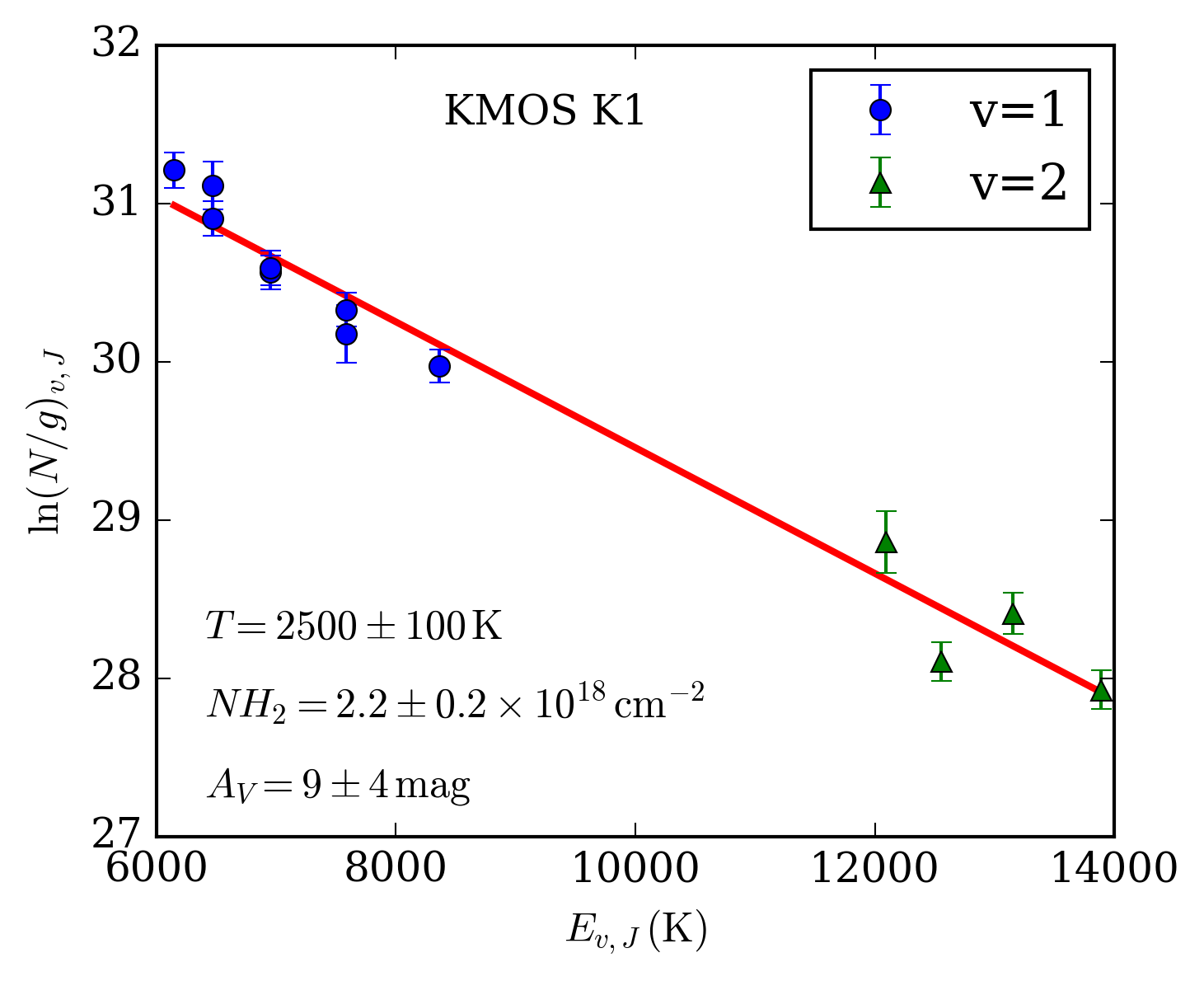}
    \caption{Ro-vibrational diagram for the integrated line fluxes of \hmol\ lines from knot K1. The plot is fitted to yield estimates of \hmol\ temperature and \hmol\ column density. The ro-vibrational diagrams for the remaining knots can be found in Appendix \ref{RV_appendix}.}
    \label{fig:eg_RV_plot}
\end{figure}

For each knot (i.e. the pixels within the 3$\sigma$ contour), we measured the integrated line fluxes of all \hmol\ lines where a Gaussian profile could be fitted with a $S/N\,\gtrsim$\,10 relative to the local continuum, and we created ro-vibrational diagrams (Boltzmann plots) using those \hmol\ line fluxes \citep[see, e.g.][]{caratti2008, caratti2015}. The \hmol\ transitions detected, their wavelengths and integrated line fluxes are given in Table \ref{tab:h2_lines}. An example of the ro-vibrational diagram of knot K1 is given if Fig. \ref{fig:eg_RV_plot}, and the remaining plots are presented in Fig. \ref{fig:all_RV_plots}. All of the plots display a thermal distribution, reinforcing the fact that we are observing shocked emission. Median \av\ values were computed for each knot location and linear regression lines were fitted to the diagrams. This yielded estimates of \hmol\ temperatures and \hmol\ column densities ranging from $\sim$\,2100\,K to $\sim$\,2800\,K, and from 10$^{18}$\,cm$^{-2}$ to 10$^{21}$\,cm$^{-2}$, respectively. The values obtained for each knot and physical property (extracted area, \av, T, N(\hmol)) are presented in Table \ref{tab:physical_properties}.

In the case of knots K2 and K3 from KMOS, we use the same \av\ value derived from the SINFONI K2* and K3 knots, respectively. As mentioned before, the KMOS data set belongs to a science verification programme and is thus not as reliable as the SINFONI observations. Given that these knots overlap (even though K2* is only part of K2), the values of \av\ derived from the SINFONI map take precedence and are used for fitting the ro-vibrational diagrams. The values of \hmol\ column densities derived for these knots are higher for SINFONI than for KMOS (see Table \ref{tab:physical_properties}), which may be due to the absolute calibrations of the data sets. Moreover, the discrepancy between K2 and K2* is further enhanced by different areas of the contours.

We also computed lower limits for the masses of the \hmol\ knots, which range from 5\ee{-4}\,\msun\ to 5\ee{-3}\,\msun. We can only report on lower limits for the \hmol\ masses because our NIR data only traces the \hmol\ warm component ($T\sim2000-4000\,$K). For additional details on the calculation of the mass limits and further discussion on the warm and cold component contributions, see Appendix \ref{mass_appendix}. Table \ref{tab:physical_properties} contains the values for \hmol\ areas (in arcsec$^2$ and in steradians) and the lower mass limits derived. Our results regarding the areas, temperatures, and column densities of knots are consistent with those of other massive jets estimated from the warm component by \cite{caratti2008,caratti2015}.

\begin{table*}
    \caption{Physical properties of knots identified in the KMOS and SINFONI data.}
    \label{tab:physical_properties}
    \centering
    \begin{tabular}{lcccccc}
    \hline \hline
    \noalign{\smallskip}
    Knot & Area & Area & \av & T & N(H$_2$) & Mass \\
    & (arcsec$^2$) & (sr) & (mag) & (K) & (cm$^{-2}$) & (M$_\odot$) \\
    \noalign{\smallskip}
    \hline
    \noalign{\smallskip}
    KMOS & & & & & & \\
    K1             & 43.7 & 1.0\ee{-9}  &  9\,$\pm$\,4           & 2500\,$\pm$\,100 & (2.2\,$\pm$\,0.2)\ee{18} & 4.8\ee{-4} \\
    K2             & 59.4 & 1.4\ee{-9}  & 19\,$\pm$\,3$^\dagger$ & 2800\,$\pm$\,300 & (8.9\,$\pm$\,1.2)\ee{18} & 2.7\ee{-3} \\
    K3             & 4.6  & 1.1\ee{-10} & 17\,$\pm$\,1$^\dagger$ & 2100\,$\pm$\,100 & (6.3\,$\pm$\,0.8)\ee{19} & 1.5\ee{-3} \\
    K4             & 11.3 & 2.6\ee{-10} & 20\,$\pm$\,4           & 2400\,$\pm$\,200 & (1.9\,$\pm$\,0.3)\ee{19} & 1.1\ee{-3} \\
    K5             & 50.5 & 1.2\ee{-9}  & 13\,$\pm$\,2           & 2400\,$\pm$\,100 & (8.9\,$\pm$\,0.8)\ee{18} & 2.3\ee{-3} \\
    K6$^\ddagger$  & 0.9  & 2.1\ee{-11} & -              & -    & -          & - \\
    K7             & 30.1 & 7.1\ee{-10} & 11\,$\pm$\,2           & 2200\,$\pm$\,100  & (9.8\,$\pm$\,0.9)\ee{18} & 1.5\ee{-3} \\
    \noalign{\smallskip}
    \hline
    \noalign{\smallskip}
    SINFONI & & & & & & \\
    K2* & 3.2 & 7.5\ee{-11} & 19\,$\pm$\,3 & 2100\,$\pm$\,200 & (2.0\,$\pm$\,0.3)\ee{20} & 3.2\ee{-3} \\
    K3  & 2.9 & 6.9\ee{-11} & 17\,$\pm$\,1 & 2200\,$\pm$\,100 & (3.1\,$\pm$\,0.2)\ee{20} & 4.7\ee{-3} \\
    K8  & 0.2 & 4.8\ee{-12} & 23\,$\pm$\,3 & 2400\,$\pm$\,100 & (3.6\,$\pm$\,0.2)\ee{21} & 3.8\ee{-3} \\
    \noalign{\smallskip}
    \hline
    \end{tabular}
    \tablefoot{The values of \hmol\ masses derived are lower limits (see text for more detail). \\
    $^\dagger$ The visual extinction derived from the SINFONI observations is more reliable than the one derived from KMOS data, which is part of a science verification programme.\\ $^\ddagger$ The location of knot K6 did not contain visual extinction pixels with enough signal-to-noise, thus we could not derive the remaining physical properties for this knot.}
\end{table*}

\subsection{CO outflows}
\label{CO_outflows}

\begin{figure*}[h!]
    \includegraphics[width=0.94\hsize]{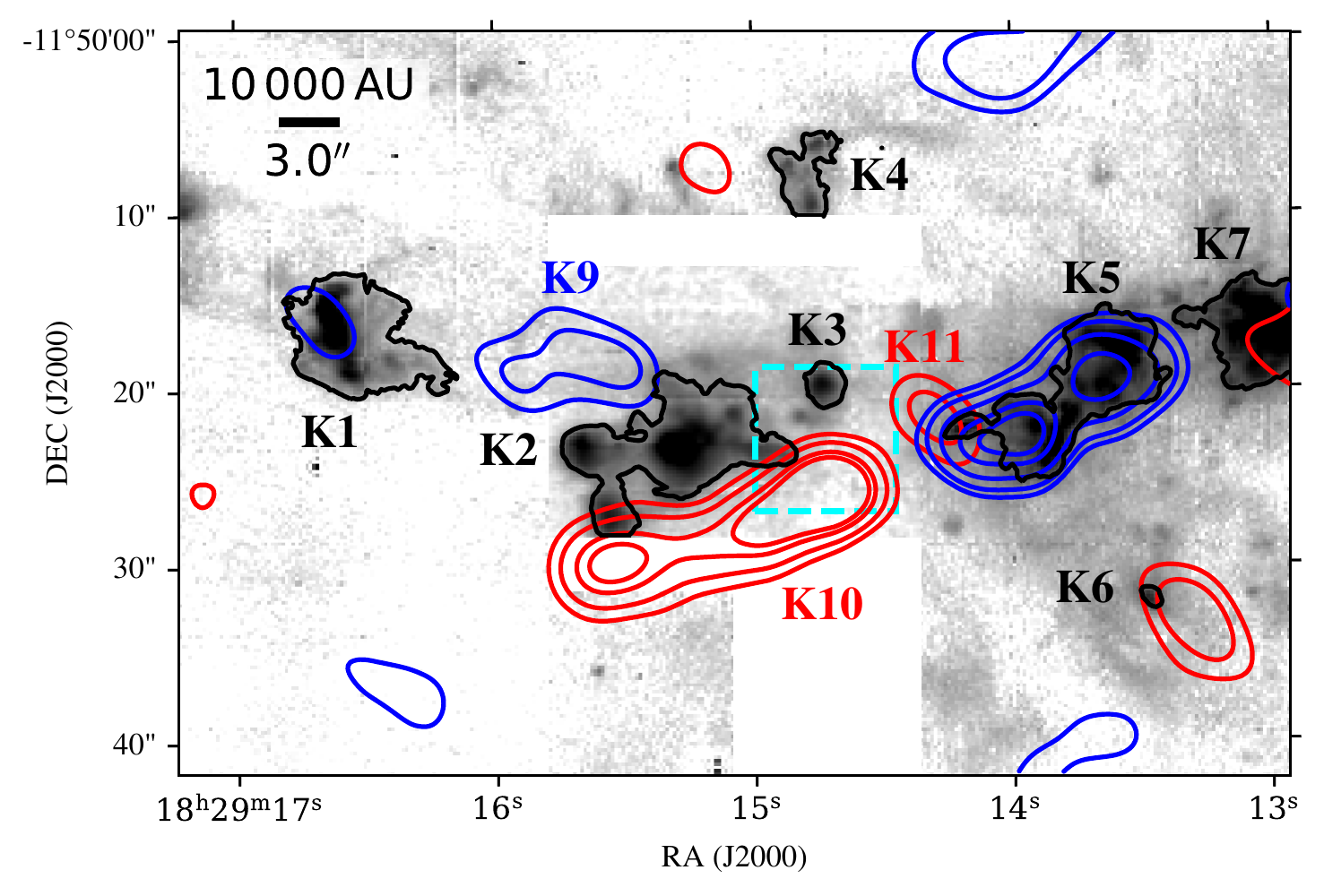}
    \caption{KMOS \hmol\ continuum-subtracted emission map with CO (2-1) outflow contours from SMA data (blue contours: 2$\sigma$, 3$\sigma$, 5$\sigma$, 8$\sigma$, 10$\sigma$, where $\sigma=0.4166$ Jy/beam; red contours: 3$\sigma$, 5$\sigma$, 8$\sigma$, 10$\sigma$, where $\sigma=0.5307$ Jy/beam). The dashed cyan rectangle defines the SINFONI field of view.}
    \label{fig:KMOS_SMA}
\end{figure*}

Figure \ref{fig:KMOS_SMA} presents the CO (2$-$1) (230.5\,GHz/1.3\,mm) emission observed with SMA, overlapped with the \hmol\ continuum-subtracted KMOS emission map. The blue wing of CO was integrated in the velocity interval [-55.7, 24.3]\,\kms, and the red wing was integrated over the range [54.3, 114.3]\,\kms (rest velocity is 43.7\,\kms). 

In the figure, we see a strong CO bipolar outflow in the SE\,-\,NW direction, where the red and blue lobes are clearly defined, extending for $\sim$\,20\arcsec and $\sim$\,15\arcsec, respectively. Another two blue-shifted knots are found to the east of the field of view, following the \hmol\ outflow, which reach a distance of $\sim$\,35\arcsec from the centre. Two more red-shifted CO knots are outlined, one in the central region (between \hmol\ knots K3 and K5), and another at the south-west corner of the image ($\sim$\,19\arcsec away from the centre). We have kept the same knot naming for the CO knots overlapping clearly with NIR ones and added new labels for those that do not have a clear correspondence with NIR structures.

The comparison with NIR data shows that the blue lobe from the CO (2$-$1) bipolar outflow is in agreement with \hmol\ knot K5 in the SE\,-\,NW outflow. The red lobe which is the counterpart of K5 (labelled as K10) lies in a region that was not covered by the NIR data set or that does not have enough signal-to-noise in the \hmol\ map. Moreover, there is a clear red-shifted CO knot in the central region (labelled as K11), which may either be part of the SE\,-\,NW outflow (K10-K5) or may be related to the red-shifted bow shock found in K7, forming another SE\,-\,NW outflow that somewhat overlaps with the K10-K5 outflow. As mentioned before, the red-shifted structures are not easily detected in our NIR observations, but they are clearly traced by the CO. 

On the other hand, regarding the NE\,-\,SW outflow, we can see that there is a marginal CO detection ($2\sigma$) in the blue-shifted \hmol\ knot K1. There is also an additional blue CO knot in this structure that is not detected in the NIR (labelled as K9). On the red-shifted south-west, we have detected a red CO knot that matches the \hmol\ knot K6. Although the signal-to-noise in this region is not enough to evidence a red-shifted \hmol\ knot to the naked eye, as discussed in Sect. \ref{velocity_maps}, the average radial velocity measured within the limits of K6 (\mbox{$\sim$\,10\,\kms}) matches the red-shifted CO contours. In summary, the red and blue outflows traced by the CO (2$-$1) data support the idea that several outflows (at least three) are launched from the \iras\ region, and our hypothesis that the blue lobes observed in NIR belong to different outflows, for which the red lobes are not observed at that wavelength. 

\subsection{Spectral energy distribution (SED)}
\label{SED}

\begin{figure}
    \centering
    \includegraphics[width=1.0\hsize]{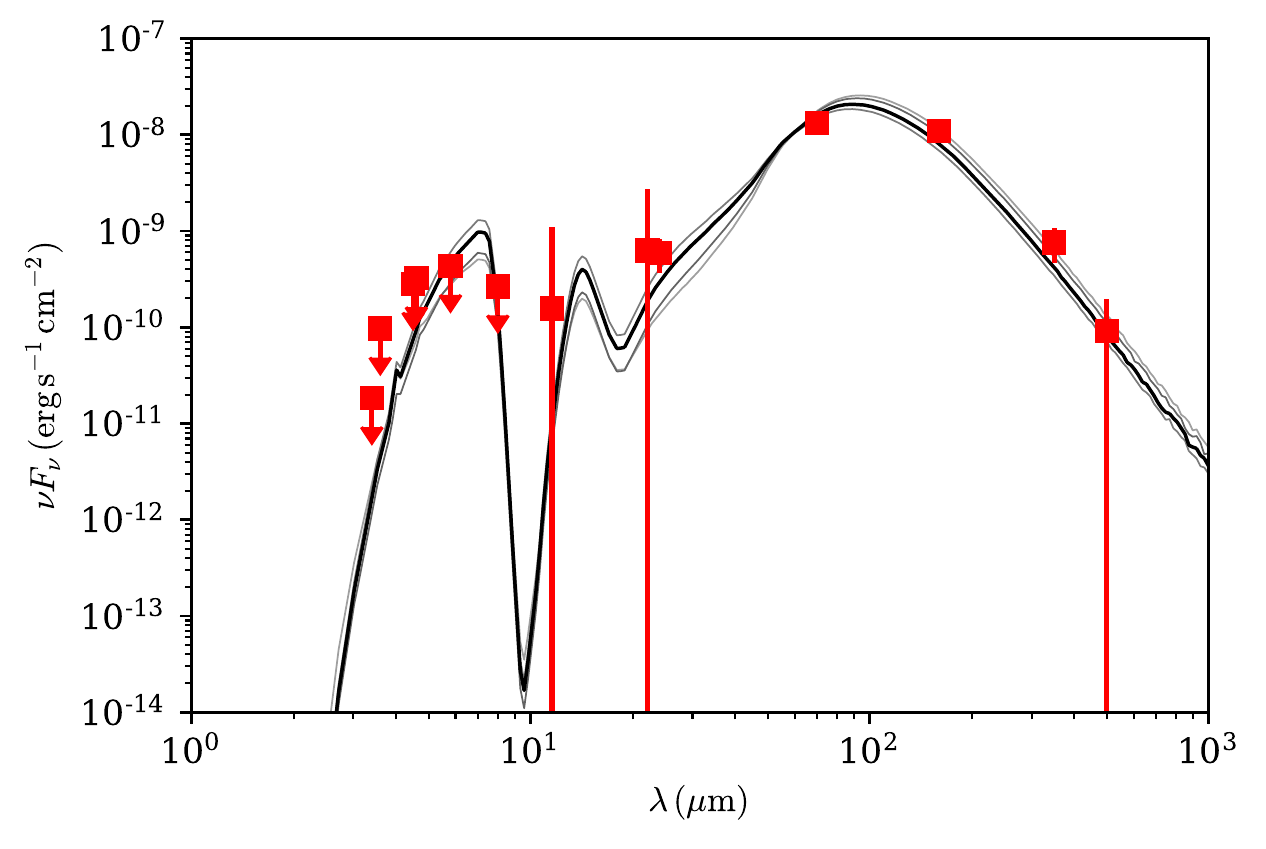}
    \caption{Spectral energy distribution (SED) model results from core accretion radiative transfer models of \cite{zhangtan2018} for \iras. The best five SED models are shown in grey scale with the best model represented with the black line (see Table\,\ref{tab:SED_parameters}) while observations are shown as red squares with the down arrow indicating upper limits (see Table\,\ref{tab:SED_flux_densities}).}
    \label{fig:SED}
\end{figure}
We built the spectral energy distribution (SED) for the \iras\ star-forming region and fitted it with \cite{zhangtan2018} protostellar radiative transfer models, following the methods described in \cite{debuizer2017, liu2019, liu2020}. Though we are likely in the presence of a protostellar cluster in \iras\ (see Sect. \ref{discussion}), here we initially assume a scenario where a single massive source dominates the luminosity of the region. This is necessitated by the fact that at Herschel wavelengths, which constrain the SED peak of the massive protostar(s), we cannot resolve individual sources in the cluster. This is also true for Spitzer data where we cannot resolve individual sources either. The approach described in the aforementioned papers has been automatised in the Python package \textit{sedcreator}\footnote{\url{https://github.com/fedriani/sedcreator}\\or \url{https://pypi.org/project/sedcreator/}} that also updates the IDL code\footnote{\url{https://zenodo.org/record/1134877#.YRJl4ZMza84}} presented in \citet{zhangtan2018}. In short, the package performs circular aperture, background-subtracted photometry on data of different wavelengths within the infrared regime, and then fits the SED with the radiative transfer models from \cite{zhangtan2018}, which are based on the Turbulent Core Accretion model \citep{mckeetan2003}. Recall that the \citet{zhangtan2018} models consider all fluxes below 8.0\,\micron\ as upper limits since these fluxes may be affected by polycyclic aromatic hydrocarbon (PAH) emission and thermal emission from grains, and these processes are not included in the model. Full details of the Python package will be given in a forthcoming paper (Fedriani et al, in prep). The models provide estimates of key protostellar properties, such as initial core mass ($M_\mathrm{c}$), environmental clump mass surface density ($\Sigma_\mathrm{cl}$), current protostellar mass ($m_*$), viewing angle with respect to the outflow axis ($\theta_\mathrm{view}$) and amount of foreground extinction ($A_\mathrm{V}$). For the \iras\ SED, we made use of data from \textit{Spitzer} (3.6 \micron, 4.5 \micron, 5.8 \micron, 8.0 \micron, 24.0 \micron), \textit{Herschel} (70.0 \micron, 160.0 \micron, 350.0 \micron, 500.0 \micron), and WISE (3.4 \micron, 4.6 \micron, 11.5 \micron, 22.0 \micron). We set our aperture size to $16\arcsec$ based on the $70\,\mu$m image and kept it constant for all wavelengths, following the fiducial method of \cite{debuizer2017,liu2019,liu2020}. The flux density values derived are presented in Table \ref{tab:SED_flux_densities}. Even though \citet{issac2020} observed the region at millimetre wavelengths with ALMA, we did not use these fluxes to fit the SED due to the flux losses in interferometric observations. Nonetheless, we compared our best SEDs with the millimetre fluxes and found that source MM2 in \citet{issac2020} is the one that is most consistent with the model flux at that wavelength (see Sect.\,\ref{discussion} for a discussion on the different sources). The resulting SED and the best fit models are shown in Fig. \ref{fig:SED}, with parameters of each model described in Table \ref{tab:SED_parameters}.

The estimated mass for the main driving source ($m_*$), according to the set of five best models, is 4-8\,\msun, with bolometric luminosity from $\sim$\,0.9 to 1.2\ee{4}\,\lsun. The predicted initial mass of the core ($M_\mathrm{c}$) where this protostar is forming is between 100 and 480 \msun. Interestingly, all five best models output a visual extinction $A_V>150$, which is consistent with the fact that the very central region between the outflows is completely obscured at NIR wavelengths. The best models output a mass accretion rate onto the disc of $\sim5\ee{-4}\,\mathrm{M_\odot\,yr^{-1}}$. At this rate a massive star, that is with $m_* \sim 8 \, M_\odot$, would have formed in $\sim 16\,000\,$yr. The half opening angle ($\theta_\mathrm{w,esc}$) is in all cases $\leq12^{\circ}$, implying collimated outflows consistent with our observations. Fig.\,\ref{fig:2D_plot} shows the $\chi^2$ distribution in the 2D parameters space $\Sigma_\mathrm{cl}-M_\mathrm{c}$ (left), $m_*-M_\mathrm{c}$ (middle) and $m_*-\Sigma_\mathrm{cl}$ (right), which shows the full constraint in the main three physical parameters for the SED fitting.

We also explore the scenario where two main sources dominate the SED, since our jet analysis points towards at least two driving sources. As we cannot resolve individual sources in the main cluster region, we simply consider that the two sources contribute 50\% to the flux at each wavelengths. We then fit the resultant SED as done above. In this way, we can estimate the properties of each individual source as if they were contributing equally to the observed density flux. The results of the best five models for a source contributing 50\% to the SED can be found in Table\,\ref{tab:SED_parameters_50PERC}. In this case, the mass of the core ranges from 60 to 200\,\msun, whereas the protostellar mass is constrained to 2-4\,\msun. The visual extinction is still above 150\,mag and the mass surface density remains somewhat on the high end going from 1.0 to 3.16\,$\mathrm{g\,cm^{-2}}$. Consequently, the bolometric luminosity is smaller than the case of one protostar dominating the SED, but the best mass accretion rates onto the disc ($\sim3\times10^{-4}\,\mathrm{\msun\,yr^{-1}}$) would form a massive protostar ($m_*>8$\,\msun) in about 25\,000\,yr.

\begin{table*}    
    \caption{Parameters of the five best fit models to the \iras\ SED.}
    \label{tab:SED_parameters}
    \centering
    \begin{tabular}{cccccccccccc}
        \hline \hline 
        \noalign{\smallskip}
        $\chi^2$ & $M_\mathrm{c}$ & $\Sigma$ & $R_\mathrm{core}$ & $m_*$ & $\theta_\mathrm{view}$ & $A_V$ & $M_\mathrm{env}$ & $\theta_\mathrm{w,esc}$ & $\dot{M}_\mathrm{disc}$ & $L_\mathrm{bol,iso}$ & $L_\mathrm{bol}$ \\
        $\mathrm{}$ & ($\mathrm{M_{\odot}}$) & ($\mathrm{g\,cm^{-2}}$) & ($\mathrm{pc}$) (\arcsec) & ($\mathrm{M_{\odot}}$) & ($\mathrm{{}^{\circ}}$) & ($\mathrm{mag}$) & ($\mathrm{M_{\odot}}$) & ($\mathrm{{}^{\circ}}$) & ($\mathrm{M_{\odot}\,yr^{-1}}$) & ($\mathrm{L_{\odot}}$) & ($\mathrm{L_{\odot}}$) \\
        \noalign{\smallskip}
        \hline
        \noalign{\smallskip}
            1.28 & 120 & 3.160 & 0.05 (3.1) & 4 & 13 & 222.87 & 113.23 & 11 & 5.7\ee{-4} & 4.0\ee{4} & 1.1\ee{4} \\
            1.51 & 160 & 3.160 & 0.05 (3.1) & 4 & 13 & 220.74 & 153.10 & 9  & 6.1\ee{-4} & 3.2\ee{4} & 1.2\ee{4} \\
            1.57 & 100 & 3.160 & 0.04 (2.5) & 4 & 13 & 224.33 & 92.43  & 12 & 5.4\ee{-4} & 4.8\ee{4} & 1.0\ee{4} \\
            1.68 & 200 & 3.160 & 0.06 (3.8) & 4 & 13 & 192.65 & 191.38 & 7  & 6.5\ee{-4} & 2.5\ee{4} & 1.2\ee{4} \\
            1.98 & 480 & 0.316 & 0.29 (18.8)& 8 & 13 & 172.32 & 461.98 & 8  & 2.0\ee{-4} & 1.9\ee{4} & 9.4\ee{3} \\
        \hline
    \end{tabular}
    \tablefoot{From left to right, the parameters are: $\chi^2$, initial core mass, mean mass surface density of the clump, initial core radius, current protostellar mass, viewing angle, foreground extinction, current envelope mass, half opening angle of outflow cavity, accretion rate from disc to protostar, isotropic bolometric luminosity, intrinsic bolometric luminosity.}
\end{table*}

\section{Unveiling a region of clustered massive star formation via outflow activity} 
\label{discussion}

\begin{figure*}[h!]
    \includegraphics[width=0.94\hsize]{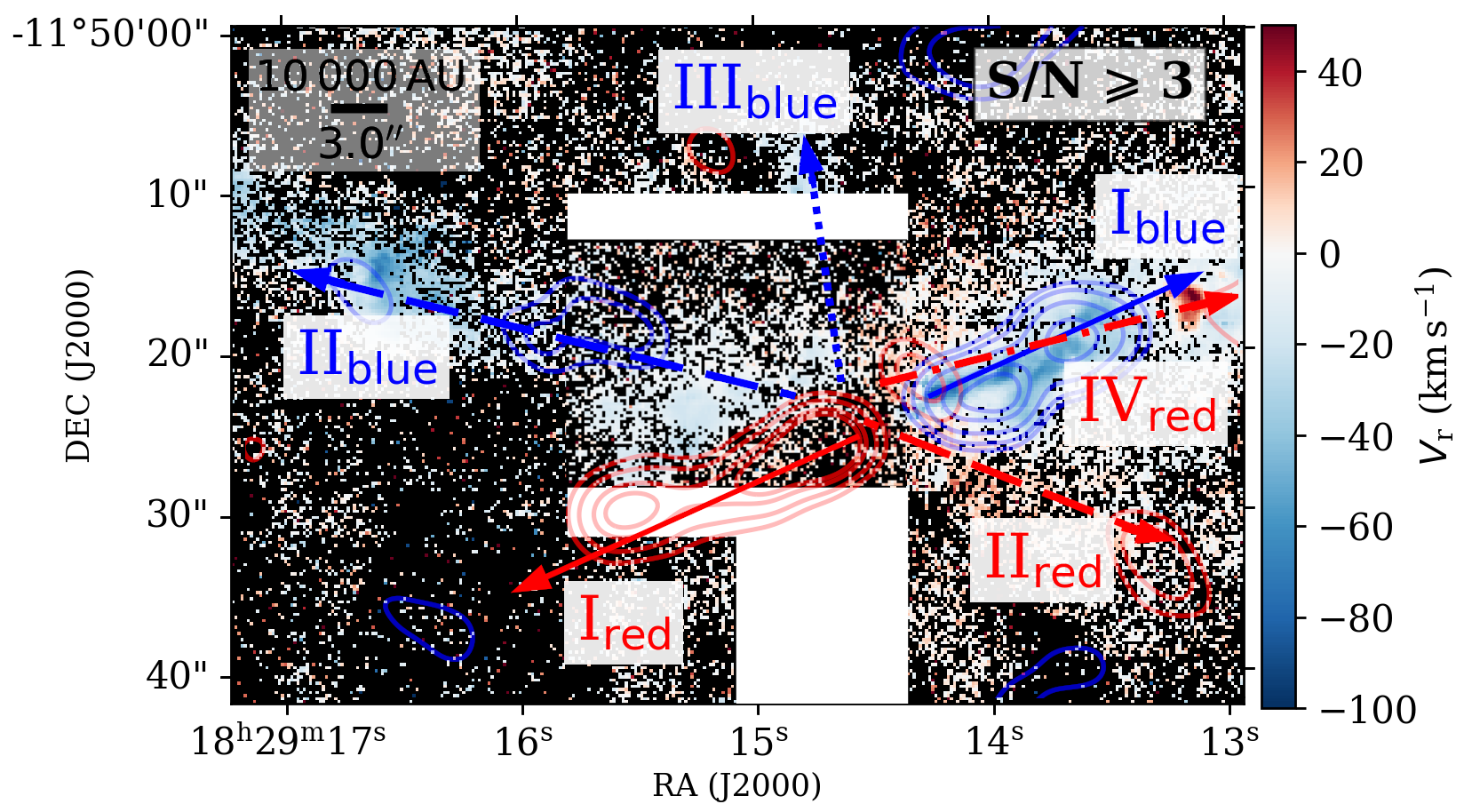}
    \caption{KMOS radial velocity map (Fig. \ref{fig:radial_velocity}) with CO (2$-$1) contours from SMA observations and possible jet and outflow morphologies represented by red and blue arrows. The dashed cyan rectangle defines the SINFONI field of view.}
    \label{fig:outflow_arrows}
\end{figure*}

\begin{figure*}
    \sidecaption
    \includegraphics[width=0.94\hsize]{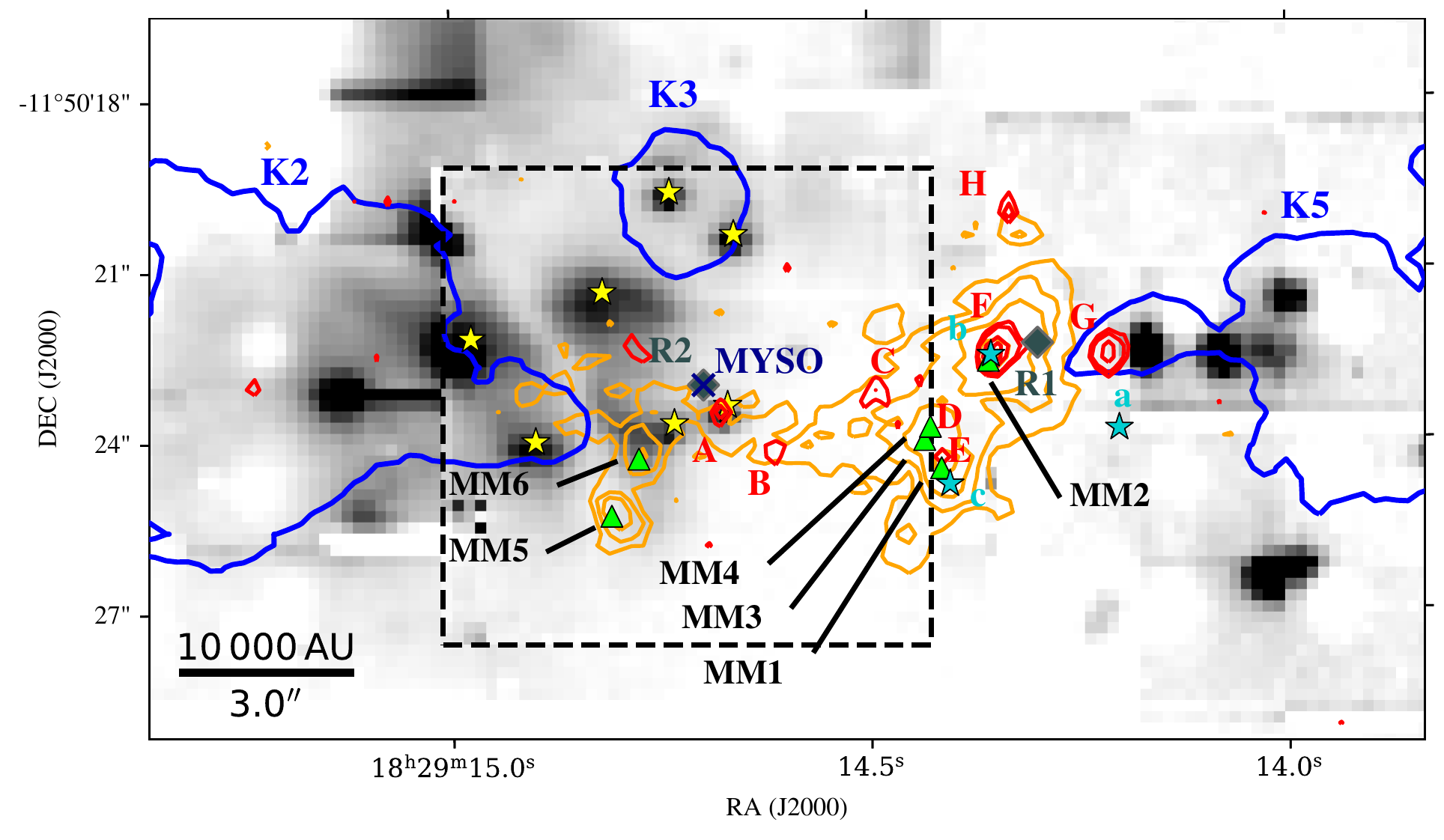}
    \caption{KMOS and SINFONI continuum emission image (SINFONI is within the dashed square, and KMOS is outside), overplotted with KMOS \hmol\ knot contours (blue, same as Fig. \ref{fig:h2_image}) and data from literature. \textit{Legend:}; red contours (A\,$-$\,H) are VLA 6\,cm data \citep{rosero2016, rosero2019}; orange contours are ALMA 2.7\,mm continuum, with green triangles (MM1\,$-$\,MM6) being the millimetre cores identified \cite{issac2020}, and the grey diamonds (R1 and R2) being uGMRT 1391.6\,MHz ionised thermal jets from the same paper; cyan stars (a, b, and c) are VLA 1.3\,cm and 7\,mm \citep{zapata2006}; dark blue cross is the location of the MYSO according to \cite{cyganowski2008}; yellow stars are the point sources identified in this work (see Sect. \ref{discussion} for further details).}
    \label{fig:literature_plots}
\end{figure*}

As seen in the previous section, the \iras\ star-forming region is nothing short of intriguing in regards to the number and location of outflows and driving sources. In Fig. \ref{fig:outflow_arrows}, we identify with red and blue arrows the different bipolar outflows found with our NIR and radio data. We report the existence of at least two bipolar outflows, extending in the SE\,-\,NW and NE\,-\,SW directions (outflows I and II). There is a possible third outflow (outflow III) as well for which we only see the blue lobe located to the north of the region. Additionally, there may be a fourth outflow (outflow IV) for which we only detect the red-shifted lobe to the west. The blue lobes of outflows I, II and III have been previously identified as MHOs 2203, 2204, and 2245, by \cite{varricatt2010} and \cite{lee2012} (see Fig. \ref{fig:h2_image}), but without any kinematic information.

The blue-shifted counterparts of outflows I and II are revealed in the \hmol\ and CO, as the emission maps show (see Figs. \ref{fig:h2_image} and \ref{fig:KMOS_SMA}), whereas the red-shifted lobes are only clearly identified in the CO. The red-shifted structures are probably very obscured in the \kband, as they have not been detected in previous NIR studies either. Furthermore, in our case, they are also partially located in regions not covered by the NIR instruments (i.e. knot K10). Our results are consistent with those of \cite{varricatt2010}, who suggested that the observed emission to the east and west could actually belong to different outflows, but could not verify this due to lack of radial velocity information. Regarding the red lobe of outflow III, it is not found in the CO data and the NIR FoV does not cover the region of its expected location. Nevertheless, the blue lobe is consistent with the NE blue outflow found by \cite{qiu2007} when tracing \mbox{SiO (2$-$1)} shocked emission. These authors also did not report a red counterpart, which could mean it is too faint to be detected or is simply not there.

The red-shifted bow shock observed to the west of the KMOS FoV is one interesting feature found in this complex, and it is difficult to explain how it fits in with the geometry of the remaining outflows. This knot is located close to the \brg\ point source S8, which could be a YSO driving the bow shock. In fact, \cite{qiu2007} also proposed that a YSO (with mass below their detection threshold) could exist close to this location, as a way to explain the several red and blue SiO knots they observed. On the other hand, \cite{issac2020} have detected C$^{17}$O (3$-$2) and C$^{18}$O (2$-$1) outflows in the SE\,-\,NW direction, with the red lobe of C$^{18}$O being located on the NW side. The emission these authors reported is consistent with knot K11 observed in our CO (2$-$1) emission map, and it is aligned with the bow shock found in the NIR knot K7. Thus we consider three possible scenarios to explain the sequence of red and blue knots along the western region: a) the large-scale blue-shifted lobe originates in the central region of \iras\ and there is a YSO close to the location of knot K7 which independently drives the red-shifted bow shock (and the blue component cannot be distinguished in our data); b) there is a precessing jet with a very large precessing angle on the plane of the sky, emitting from the central region and creating a chain of red and blue knots to the west (although in this case, we should detect a similar pattern towards the east flow, which we do not); or c) there are two distinct bipolar outflows being driven in the SE\,-\,NW direction by two different sources, which visually overlap along the line of sight; one source drives SE red-shifted emission and NW blue-shifted emission, whereas the other drives SE blue-shifted and NW red-shifted lobes. Further observations are required to better understand the origin of these knots and where their driving sources lie.

To explain the existence of multiple large-scale outflows throughout this region, we propose that \iras\ is most likely harbouring a protostellar cluster, where different sources are driving each set of outflows. Indeed, the geometry of the outflows indicates that most are launched from the same region harbouring \iras. To better understand the morphology and dynamics of this star-forming region, we plot KMOS and SINFONI emission maps with other millimetre and radio data sets and results from the literature in Fig. \ref{fig:literature_plots}.

We find that the VLA knot G \citep{rosero2016} overlaps with our knot K5 seen in KMOS and in SMA data. \cite{rosero2016} calculated a spectral index of $-0.2$ for this source, which together with the lack of association with a millimetre core, make it likely that this is a radio jet knot. Our observations of both \hmol\ and CO jet knots at those coordinates further support this classification. From this same plot of Fig. \ref{fig:literature_plots}, we can also see that sources R1, MYSO, and A are spatially coincident with point source S4 from our work. S4 has a rising continuum in the \kband, so the correspondence to compact cores and thermal jets from the literature lends support to our claim that S4 is a YSO. We note that the markers are not perfectly overlapping, but are within 1\arcsec of each other. The discrepancy could be attributed to astrometric errors in the data sets.

Further away from the main cluster region ($\sim$\,20\arcsec), towards the west, we find two point sources (i.e. S8 and S9). Both sources are \brg\ emitters, signalling youth, and lie along main outflow I, possibly contributing to the complex picture of the multiple outflows. Besides this, there is a VLA compact source in the proximity of S9, further suggesting their relationship with the main clustered region. However, it should be noted that the spectral index of this knot is $<-0.8$ \citep[labelled G19.883-0.528 in][]{rosero2016}, therefore it does not seem to be a YSO \textit{per se}, but more likely a synchrotron jet.

The sources R1/MM2/b/F (hereafter R1, RA(J2000)\,=\,18:29:14.36, Dec(J2000)\,=\,-11:50:22.5)\footnote{Coordinates are from VLA 4.9\,GHz (6cm) from \citet{rosero2016}} and R2/MYSO/A/S4 (hereafter R2, RA(J2000)=18:29:14.69, Dec(J2000)=-11:50:23.6)\footnotemark[\value{footnote}], as labelled in Fig. \ref{fig:literature_plots}, were pointed out as the main driving sources in this region \citep{issac2020}. In fact, their spectral indices are 0.6 and 0.5, respectively \citep{rosero2016}, which together with the association to dust continuum, indicates their YSO nature. Source R1 is believed to be the most massive source of the cluster \citep{rosero2016, issac2020}, and likely powering the CO \mbox{SE (blue)\,-\,NW (red)} outflow. This possibly matches our red-shifted outflow IV, however we cannot detect any NIR counterpart of the source in our data. A signature of protostellar infall was found towards the source MM6 \citep{issac2020}, which is less massive than either R1 or R2. If it is the case that R1 drives outflow IV, then R2 could be driving either outflow I or II, which are more massive, and MM6 could be powering outflow III. We note that MM6 is close to our source S6, which has \brg\ emission. So even though we do not directly detect MM6, we could be observing its light being reflected on the outflow cavity wall in source S6. However, we emphasise that this is a somewhat speculative interpretation and with the current data we cannot determine conclusively which sources drive which outflows.

The multi-wavelength review and analysis carried out by \cite{issac2020} concludes that there are possibly up to eleven sources clustered in the central region of \iras\ (we stress that some sources have multiple names, e.g. R1/MM2/b/F). Furthermore, we consider the possibility that the \brg-line emitting sources observed in the SINFONI and KMOS data are also part of the cluster, raising the total membership number to fourteen young objects in a box of $10\arcsec\times10\arcsec$ (0.0256\,pc$^2$) centred at the cluster (and to sixteen sources overall). We reiterate that the contribution of these objects to the overall outflow picture is still uncertain. Nonetheless, we provide a lower limit for the number of sources in this region. This would imply a lower limit for the stellar number density of about $14/0.1503^3\sim$\,4000\,pc$^{-3}$, assuming a uniform distribution in a cube in the very central region of the cluster. This stellar density is somewhat smaller than what is found in the central regions of some young clusters, such as $\sim$\,10$^4$\,pc$^{-3}$ in the Orion Nebula Cluster \citep{hillenbrand1998} or $\sim$\,10$^5$\,pc$^{-3}$ in R136 in 30 Doradus \citep{selman2013}, see \citet{portegies2010} for a review of young clusters. \citet{bonnell1998} proposed that, with stellar densities of about $10^8$\,pc$^{-3}$, stellar collisions could occur to explain the formation of massive protostars, and this number could be lowered to $10^6$\,pc$^{-3}$ if all massive stars are hard binaries \citep{bonnell2005}. At any event, our estimates are far from the stellar density required by these previous works.

To conclude, we find that \iras\ is a region of clustered massive star formation containing multiple collimated bipolar outflows. This suggests that massive stars can form in a relatively ordered manner even in such clustered regions. This conclusion is supported by our estimates of YSO number densities in the region, which, while uncertain, are much lower than the values theorised to be necessary for protostellar collisions to be important. While competitive accretion models do not require as high a level of stellar density, they are generally thought to lead to more disordered accretion disc and outflow orientations and symmetries \citep[see, e.g.][]{tan2014}. The results we have presented here are important constraints for such models.

\section{Summary and conclusions}
\label{conclusions}

We have studied the morphology and composition of the \iras\ massive star-forming complex via integral field unit observations from SINFONI and KMOS in the \kband, as well as CO (2$-$1) data from SMA. Our results can be summarised as follows:

\begin{enumerate}

\item In the SINFONI field of view we observed three point sources (from S5 to S7) emitting \brg, which is consistent with YSO activity. Furthermore, we report another source (S4) that displays a rising \kband\ continuum and that has shown other YSO characteristics in previous studies \citep{rosero2016, issac2020}. Further away from the central cluster, through the KMOS maps, we found two more \brg\ emitting sources (S8 and S9), but it is undetermined whether they belong to the \iras\ central cluster. We thus proposed five new YSOs for this region (S5 to S9) based on our work, and presented further evidence to justify the youth of S4. 

\item We focussed our near-infrared analysis on the \hmol\ jet knots found in the region. We identified seven \hmol\ jet knots in the KMOS data (labelled K1 to K7) and three jet knots in the SINFONI data (K2*, K3, and K8), where two of which are the same as those resolved in KMOS. Large-scale outflow lobes are found in the directions of NE (K1, K2, K3?) and NW (K5, K7), with two smaller jet knots to the north (K4) and south-west (K6). Our observations are in good agreement with the previous NIR works that found a large-scale east-west outflow with other non-aligned knots \citep[e.g.][although with no kinematic information in the NIR]{varricatt2010, lee2012, issac2020}. 

\item We found that the \hmol\ jet emission is shock driven by computing the flux ratio of the \hmol\ transitions \mbox{1\,$-$\,0 S(1)} and \mbox{2\,$-$\,1 S(1)}. 

\item Upon inspection of the radio velocity maps, we found that the large-scale jets detected are blue-shifted, as well as the northern knot, whereas the south-west knot is red-shifted, and there is a red-shifted bow shock knot at the west edge of the KMOS field of view. It seems that the blue lobes belong to different outflows, and their red counterparts are most likely obscured as they are not clearly detected in our \kband\ data. 

\item We computed visual extinction estimates for the region, ranging from $\sim$\,10 to $\sim$\,20 magnitudes, and measured the line fluxes of all \hmol\ transitions present in each knot. With this information, we constructed and fitted ro-vibrational diagrams to yield estimates of \hmol\ temperatures and \hmol\ column densities of 2100\,K to 2800\,K and 10$^{18}$ to 10$^{21}$\,cm$^{-2}$, respectively. The areas of the \hmol\ knots vary between 0.2\arcsec and 60\arcsec, and the lower limits for the \hmol\ masses (due only to \hmol\ warm component) range from 5\ee{-4}\,\msun\ to 5\ee{-3}\,\msun. 

\item The CO (2$-$1) emission traces two clear bipolar outflows in the NE\,-\,SW and SE\,-\,NW directions, which is consistent with the works of \cite{beuther2002b, issac2020}. In general, the blue lobes of the CO outflows are in good agreement with the blue lobes of the \hmol\ jets. The red-shifted counterpart of the SE\,-\,NW CO outflow (K10) lies in a region that was not covered by the NIR observations or that has very low signal-to-noise, thus we cannot detect a corresponding \hmol\ jet. On the other hand, the red lobe of the NE\,-\,SW CO outflow overlaps with a small \hmol\ red-shifted knot (K6), suggesting that it might be the counterpart of the blue lobe to the north-east. Additionally, a red-shifted CO knot was found near the central region, just off the west of the SINFONI field of view. This knot is consistent with the \mbox{C$^{18}$O (1$-$0)} and \mbox{C$^{17}$O (3$-$2)} observations of \cite{issac2020} where a collimated SE\,-\,NW CO outflow was found, having the red-shifted emission on the NW side. Moreover, this knot is radially aligned with the \hmol\ red-shifted bow shock in K7, possibly unveiling another outflow.

\item We created the spectral energy distribution plot for \iras\ with archival data from \textit{Spitzer}, \textit{Herschel}, and WISE, and fitted it using the \cite{zhangtan2018} radiative transfer models based on the Turbulent Core Accretion model \citep{mckeetan2003}. Though this complex likely harbours a protocluster, we assume the scenario where one massive driving source dominates the luminosity of the region. The five best fit models provided estimates for the mass of the driving source, bolometric luminosity, predicted initial mass of the core, extinction, and mass accretion rate of 4\,-\,8\,\msun, $\sim$\,$0.9-1.2\ee{4}$\lsun, 100\,-\,480\,\msun, $>$\,$150$\,mag, and $\sim$\,$2.0-6.5\ee{-4}$\,\msun\,yr$^{-1}$, respectively. We also explored the scenario where two sources dominate the SED equally. We then fitted an SED with 50\% flux at each wavelength to represent the individual contribution of each source. Consequently, we obtained smaller values for the central mass of the protostar that ranges from 2 to 4\,\msun forming in a core with initial mass of 60-200\,\msun.

\end{enumerate}

Our main conclusion of this study regards the outflow geometry in the \iras\ region, for which we report at least two, and up to four different outflows. Both the \hmol\ and CO (2$-$1) observations trace two large-scale bipolar outflows in the SE\,-\,NW and NE\,-\,SW direction (outflows I and II), and a possible red-shifted outflow to the west (outflow IV), whereas the possible blue-shifted north outflow (outflow III) is only revealed in the \hmol\ maps. We compare our outflows with the locations of \iras\ massive sources and thermal jets listed in the literature. The main driving sources of the cluster are believed to be sources R2/MYSO/A/S4 (RA(J2000)\,=\,18:29:14.69, Dec(J2000)\,=\,-11:50:23.6) and R1/MM2/b/F (RA(J2000)\,=\,18:29:14.36, Dec(J2000)\,=\,-11:50:22.5).

Further observations at high angular resolution in the MIR (e.g. with SOFIA) are required to find the counterparts of the outflows presented in this study. Moreover, such data will help us to clearly assess the origin of the red-shifted bow shock, and to check whether point sources S8 and S9 truly belong to the \iras\ cluster. 

In conclusion, our study strongly supports the scenario of several outflows driven by multiple sources and demonstrates that NIR observations have high diagnostic power in probing high-mass star-forming regions. The outflows show high degrees of collimation on large scales suggesting that massive stars can form in a relatively ordered manner, consistent with core accretion models, even in a clustered region.

\begin{acknowledgements}
We thank the referee for their comments and suggestions which improved the clarity of the manuscript. We would like to thank Prof. Dr. Henrik Beuther for assistance with the SMA observations. A.R.C.S. acknowledges funding from Chalmers Astrophysics and Space Sciences Summer (CASSUM) research programme. R.F. acknowledges funding from the European Union’s Horizon 2020 research and innovation programme under the Marie Sklodowska-Curie grant agreement No 101032092. A.C.G. has been supported by PRIN-INAF-MAIN-STREAM 2017 ``Protoplanetary disks seen through the eyes of new-generation instruments'' and by PRIN-INAF 2019 ``Spectroscopically tracing the disk dispersal evolution (STRADE)''. J.C.T. acknowledges support from ERC grant MSTAR, VR grant 2017-04522, and NSF grant 1910675. G.C and P.G acknowledge support from Chalmers Initiative Cosmic Origins (CICO) postdoctoral fellowships. This research made use of Photutils, version 1.0.0, an Astropy \citep{astropy2013,astropy2018} package for detection and photometry of astronomical sources \citep{larry_bradley_2020_4044744}.
\end{acknowledgements}

\bibliographystyle{aa}
\bibliography{biblio.bib}

\begin{appendix}

\section{KMOS \kband\ continuum image}
\label{KMOS_continuum_appendix}

\begin{figure*}[h]
    \centering
    \includegraphics[width=0.94\hsize]{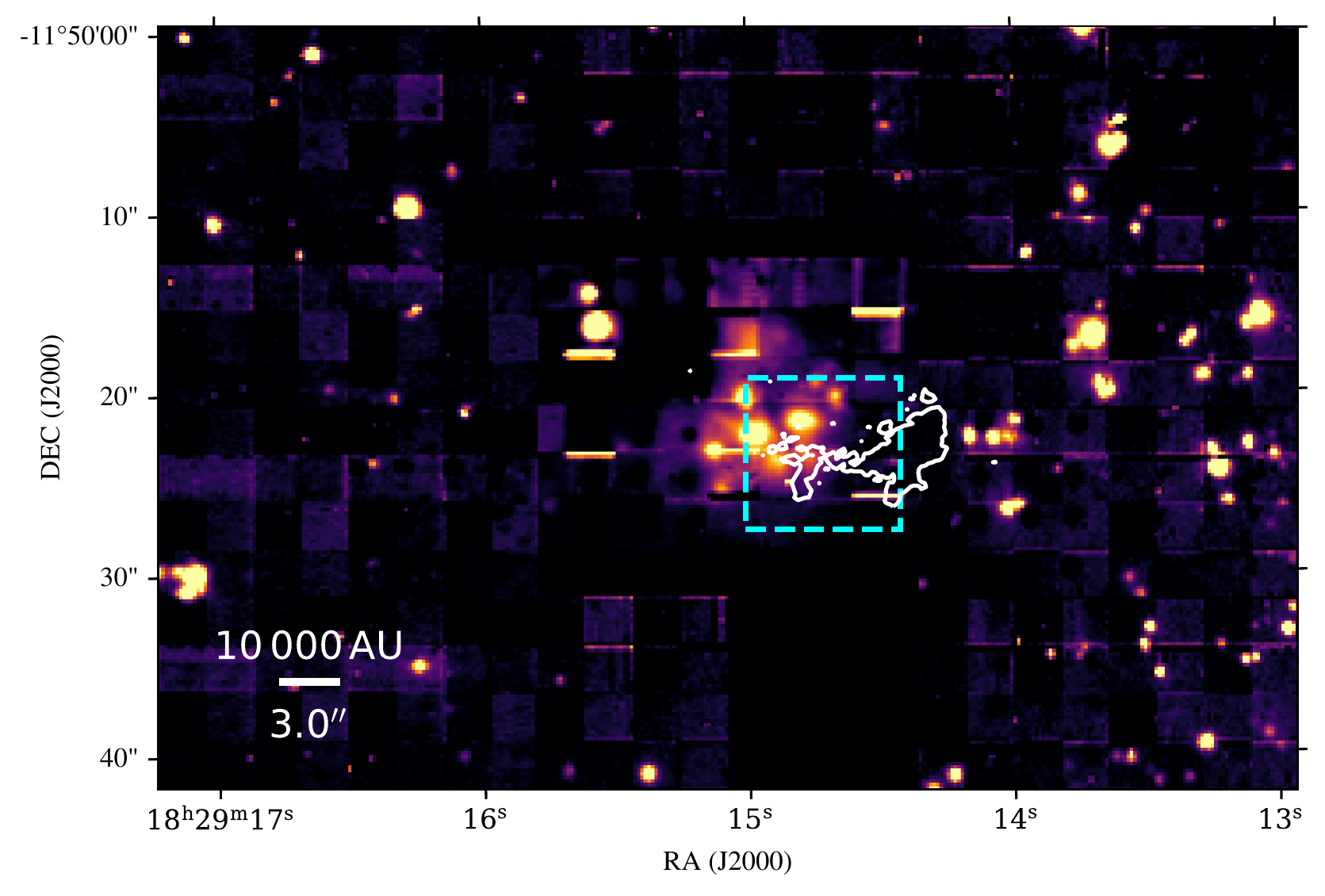}
    \caption{KMOS \kband\ continuum image (313 line-free channels across the entire \kband). The horizontal lines seen are due to poor data reduction, as they coincide with the edges of the individual fields of view of the IFUs. The dashed cyan rectangle defines the SINFONI field of view; the white contours are the ALMA 2.7\,mm data analysed in \cite{issac2020} (see text for more detail).}
    \label{fig:kmos_continuum}
\end{figure*}

\section{Integrated line fluxes of knots}
\label{line_fluxes_appendix}

\begin{sidewaystable*}[h]
    \label{tab:h2_lines}
    \caption{Integrated line fluxes of \hmol\ lines detected in each jet knot.}
    \centering
    \begin{tabular}{l c|c c c c c c c|c c c}
    \hline\noalign{\smallskip}\hline
    \noalign{\smallskip}
    \noalign{\smallskip}
    Transition & Wavelength & \multicolumn{10}{c}{Integrated line fluxes (10$^{-15}$ erg s$^{-1}$ cm$^{-2}$)} \\
    & in vacuum & \multicolumn{7}{c}{KMOS} & \multicolumn{3}{c}{SINFONI} \\
    & (\micron) & K1 & K2 & K3 & K4 & K5 & K6 & K7 & K2* & K3 & K8 \\
    \noalign{\smallskip}
    \hline
    \noalign{\smallskip}
    1\,$-$\,0 S(3) & 1.9576 &  3.93$\pm$0.42  & 4.42$\pm$0.46  &  3.71$\pm$0.42  &  2.71$\pm$0.31  &  12.4$\pm$1.28 &  -              &  12.1$\pm$1.27  &  15.5$\pm$1.70  &  24.1$\pm$2.52  &  8.72$\pm$1.22 \\
    1\,$-$\,0 S(2) & 2.0338 &  1.50$\pm$0.16  & 3.20$\pm$0.34  &  2.65$\pm$0.30  &  1.13$\pm$0.14  &  5.02$\pm$0.52 &  1.75$\pm$0.21  &  4.14$\pm$0.43  &  5.14$\pm$0.60  &  6.74$\pm$0.74  &  3.03$\pm$0.43 \\
    2\,$-$\,1 S(3) & 2.0735 &  0.73$\pm$0.09  & 1.30$\pm$0.15  &  0.77$\pm$0.12  &  0.41$\pm$0.06  &  1.96$\pm$0.21 &  0.64$\pm$0.11  &  1.54$\pm$0.17  &  1.57$\pm$0.27  &  2.29$\pm$0.28  &  1.42$\pm$0.26 \\
    1\,$-$\,0 S(1) & 2.1218 &  3.99$\pm$0.43  & 8.40$\pm$0.90  &  7.74$\pm$0.83  &  3.11$\pm$0.33  &  14.8$\pm$1.52 &  5.55$\pm$0.60  &  12.4$\pm$1.28  &  13.2$\pm$1.41  &  19.1$\pm$1.96  &  9.17$\pm$1.02 \\
    2\,$-$\,1 S(2) & 2.1542 &  0.32$\pm$0.04  & 0.81$\pm$0.13  &  -              &  0.35$\pm$0.05  &  0.98$\pm$0.12 &  0.38$\pm$0.08  &  0.72$\pm$0.09  &  0.75$\pm$0.21  &  1.14$\pm$0.20  &  0.44$\pm$0.10 \\
    1\,$-$\,0 S(0) & 2.2233 &  1.00$\pm$0.11  & 2.15$\pm$0.23  &  2.16$\pm$0.25  &  0.83$\pm$0.10  &  3.83$\pm$0.39 &  1.55$\pm$0.20  &  2.97$\pm$0.31  &  2.53$\pm$0.36  &  5.14$\pm$0.57  &  2.10$\pm$0.35 \\
    2\,$-$\,1 S(1) & 2.2477 &  0.51$\pm$0.06  & 0.95$\pm$0.11  &  0.83$\pm$0.12  &  0.46$\pm$0.07  &  1.76$\pm$0.19 &  0.71$\pm$0.11  &  1.17$\pm$0.13  &  1.05$\pm$0.22  &  2.43$\pm$0.30  &  -             \\
    2\,$-$\,1 S(0) & 2.3556 &  0.19$\pm$0.04  & 0.65$\pm$0.11  &  -              &  -              &  0.68$\pm$0.09 &  -              &  0.42$\pm$0.06  &  -              &  0.65$\pm$0.13  &  -             \\
    1\,$-$\,0 Q(1) & 2.4066 &  4.30$\pm$0.48  & 6.66$\pm$0.71  &  6.48$\pm$0.72  &  4.49$\pm$0.51  &  15.7$\pm$1.65 &  2.29$\pm$0.46  &  13.3$\pm$1.41  &  14.8$\pm$1.67  &  23.1$\pm$2.44  &  11.5$\pm$1.49 \\
    1\,$-$\,0 Q(2) & 2.4134 &  1.53$\pm$0.23  & 2.65$\pm$0.31  &  2.85$\pm$0.36  &  1.83$\pm$0.25  &  5.36$\pm$0.61 &  -              &  4.68$\pm$0.57  &  3.83$\pm$0.61  &  6.91$\pm$0.82  &  3.20$\pm$0.55 \\
    1\,$-$\,0 Q(3) & 2.4237 &  3.52$\pm$0.39  & 5.77$\pm$0.62  &  5.97$\pm$0.67  &  3.37$\pm$0.38  &  13.0$\pm$1.36 &  1.48$\pm$0.18  &  10.3$\pm$1.08  &  13.1$\pm$1.53  &  19.7$\pm$2.11  &  10.9$\pm$1.51 \\
    1\,$-$\,0 Q(4) & 2.4375 &  0.95$\pm$0.17  & 1.91$\pm$0.24  &  1.78$\pm$0.26  &  1.29$\pm$0.20  &  5.46$\pm$0.66 &  -              &  3.81$\pm$0.53  &  3.22$\pm$0.50  &  5.86$\pm$0.71  &  3.67$\pm$0.66 \\
    \noalign{\smallskip}
    \hline
\end{tabular}
\tablefoot{The dashes ('-') stand for non detections or the Gaussian fit returned $S/N\,<$\,10.}
\end{sidewaystable*}

\section{Methods}

As is demonstrated in the main text, the available NIR data allow us to infer a number of crucial parameters by using \hmol\ emission lines. In particular, we can estimate the visual extinction, for which the process is detailed in Appendix \ref{extinction_appendix}. Furthermore, from the column density estimations retrieved from the ro-vibrational diagrams, the \hmol\ mass of the knots can be estimated as explained in Appendix \ref{mass_appendix}.

\subsection{Extinction calculation}
\label{extinction_appendix}

By choosing two lines that originate from the same upper ro-vibrational level (in this case, \hmol\ \mbox{1\,$-$\,0 S(1)} and \hmol\ \mbox{1\,$-$\,0 Q(3)}), the theoretical flux ratios are simplified, depending only on the Einstein Coefficients (A$_\mathrm{1}$, A$_\mathrm{2}$, i.e. the transition probabilities) and the wavelengths of the transitions. Thus the colour excess, $E(\lambda_\mathrm{1}-\lambda_\mathrm{2})$, can be found via the following equation:
\begin{equation}
\label{eq:fluxratio}
    \frac{F_{\lambda_\mathrm{1}}}{F_{\lambda_\mathrm{2}}} = \frac{\mathrm{A_1}\lambda_2}{\mathrm{A_2}\lambda_1}\,10^{-E(\lambda_\mathrm{1}-\lambda_\mathrm{2})/2.5},
\end{equation}
The colour excess is then related to \av\ by:
\begin{equation}
\label{eq:colourexcess_av}
    E(\lambda_\mathrm{1}-\lambda_\mathrm{2}) = A_\mathrm{V}\, \alpha\, (\lambda_1^\beta - \lambda_2^\beta),
\end{equation}
where $\alpha$ and $\beta$ depend on the chosen extinction law. In this work, we adopt the extinction law by \cite{rieke1985} ($\alpha=0.42,\,\beta=-1.75$). Combining Eqs. \ref{eq:fluxratio} and \ref{eq:colourexcess_av}, and solving for the visual extinction, we find that:
\begin{equation}
\label{eq:av}
    A_\mathrm{V} = \log_{10} \left( \frac{F_{\lambda_\mathrm{1}}}{F_{\lambda_\mathrm{2}}} \frac{\lambda_2 \mathrm{A_1}}{\lambda_1 \mathrm{A_2}} \right) \left( -\frac{2.5}{0.42(\lambda_1^{-1.75} - \lambda_2^{-1.75})} \right).
\end{equation}

It is important to note that the \av\ values derived in the maps are to be treated as approximate estimates for two main reasons: 1) the \mbox{1\,$-$\,0 Q(3)} line is located in a very noisy atmospheric transmission region, and 2) the chosen transitions are located only $\sim$\,0.3 \micron\ apart, which is a much shorter separation than would be desirable for this method \citep[see][]{nisini2008}. Alternatively, the visual extinction could have been estimated using another pair of \hmol\ lines that fulfils the criterion \mbox{1\,$-$\,0 S($i$)}/\mbox{1\,$-$\,0 Q($i+2$)}, such as the pairs \mbox{1\,$-$\,0 S(0)}/\mbox{1\,$-$\,0 Q(2)} or \mbox{1\,$-$\,0 S(2)}/\mbox{1\,$-$\,0 Q(4)}. However, these lines have even lower signal-to-noise ratios than those of our chosen pair. Thus using the line fluxes of 1\,$-$\,0 S(1)/1\,$-$\,0 Q(3) is the best way to obtain visual extinction estimates in the jet regions with the available data.

\subsection{\hmol\ mass calculation}
\label{mass_appendix}

We computed lower limits for the masses of the knots, using the following equation \citep[see, e.g.][]{caratti2008}:
\begin{equation}
    M = 2\,\mu\,m_\mathrm{H}\,N(\mathrm{H_2})\,A_\mathrm{knot}
\end{equation}
where $\mu=1.24$ is the mean atomic weight, $m_\mathrm{H}$ is the mass of the hydrogen atom, $N(\mathrm{H_2}$) is the column density, and $A_\mathrm{knot}$ is the \hmol\ area of the knot.

We can only report on lower limits for the \hmol\ masses because our NIR data only trace the \hmol\ warm component ($T\sim2000-4000\,$K). To obtain the full value of \hmol\ mass for each jet, we would require data in the mid-infrared (MIR) tracing the cold component ($T\sim500-2000\,$K) measured with the pure rotational lines between 5\,\micron\ and 28\,\micron\ \citep{caratti2008}. Table \ref{tab:physical_properties} contains the values for \hmol\ areas (in arcsec$^2$ and in steradians) and the lower mass limits derived. We find that our values for mass are about one to two orders of magnitude smaller than those presented in \cite{caratti2008,caratti2015}. This is still consistent given that, in the aforementioned works, the authors also included the contribution of the cold component from MIR data to estimate the overall knot mass. Typically, the column densities derived from the \hmol\ cold component (and thus the inferred mass of the knot) are orders of magnitude higher than those inferred with the warm and hot components in the NIR \citep[see, e.g. Fig. 1 of][]{caratti2008}. Nonetheless, our estimates for the warm component of the column density are still consistent with previous works.

\section{Ro-vibrational diagrams}
\label{RV_appendix}

\begin{figure*}[h]
    \centering
    \includegraphics[width=0.32\hsize]{referee_plots/KMOS_RV_K1.png}
    \includegraphics[width=0.32\hsize]{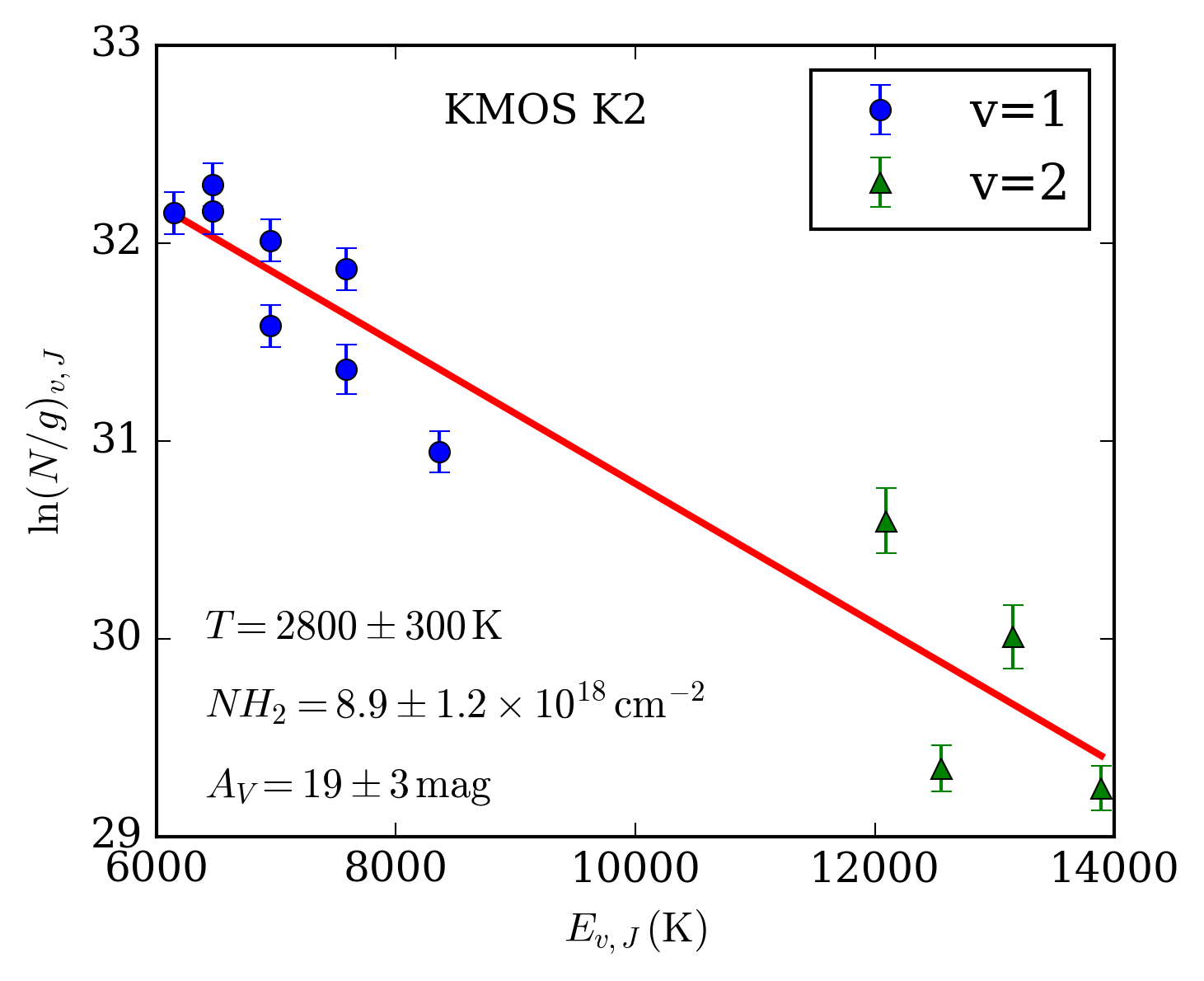}
    \includegraphics[width=0.32\hsize]{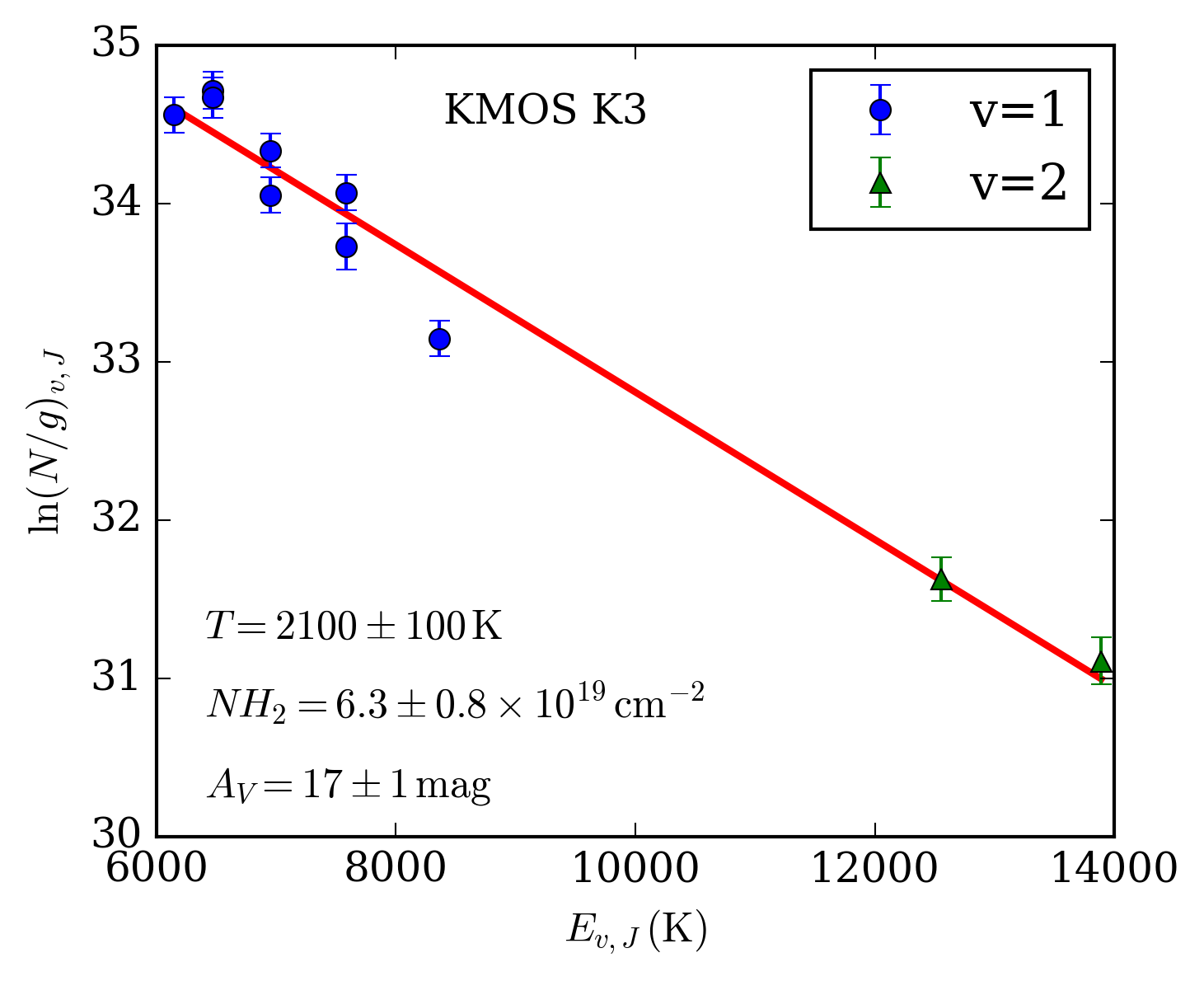}
    \includegraphics[width=0.32\hsize]{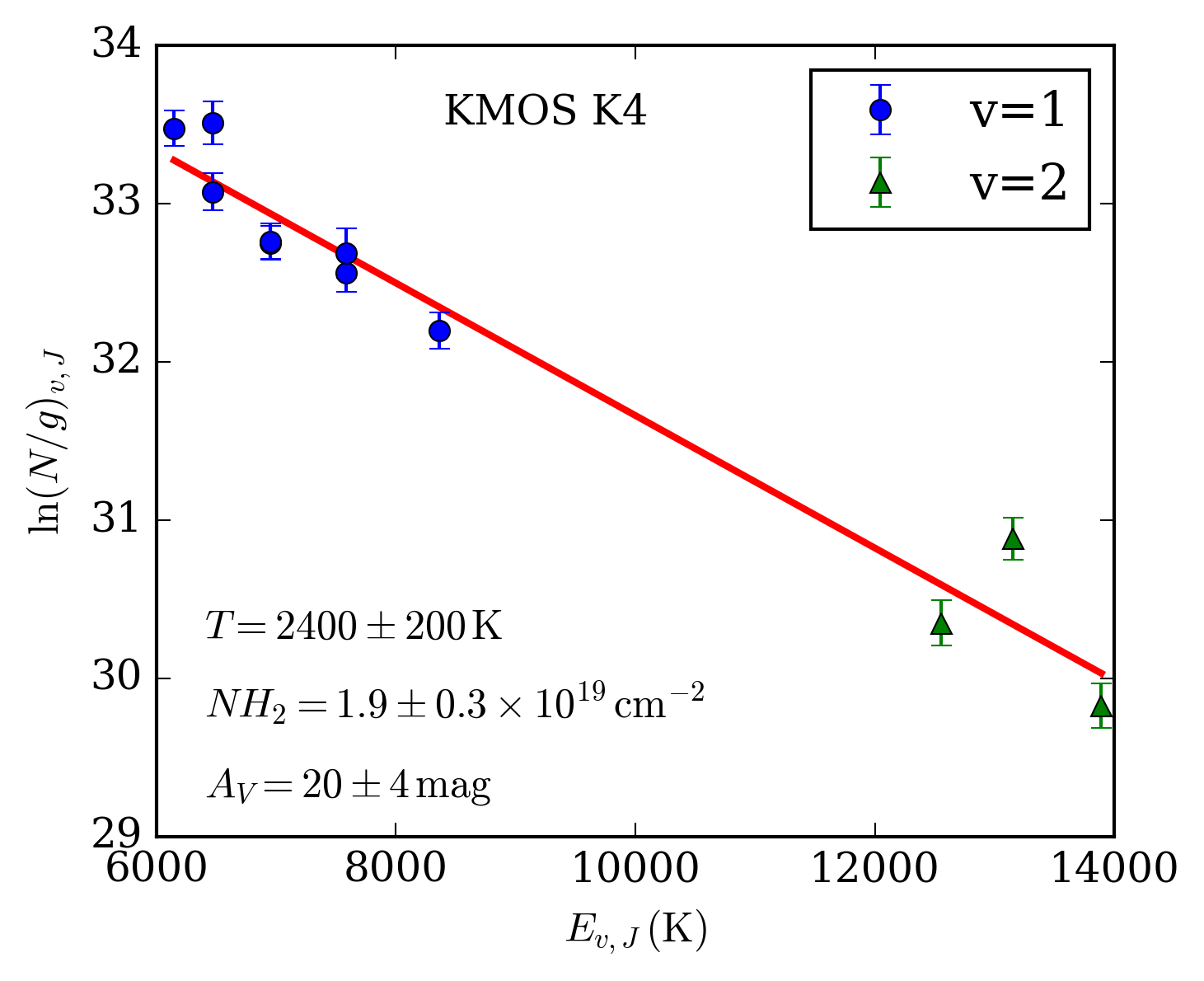}
    \includegraphics[width=0.32\hsize]{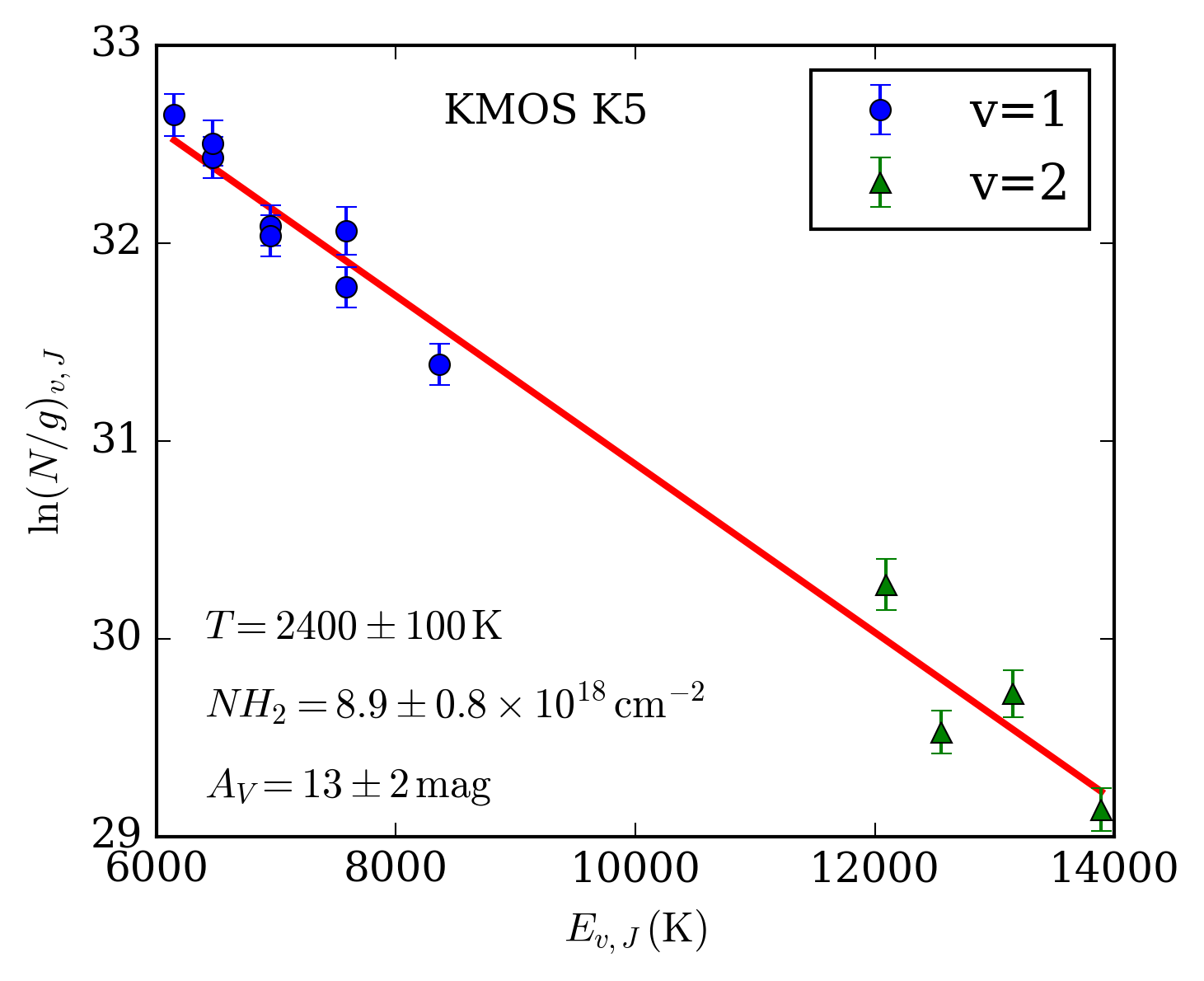}
    \includegraphics[width=0.32\hsize]{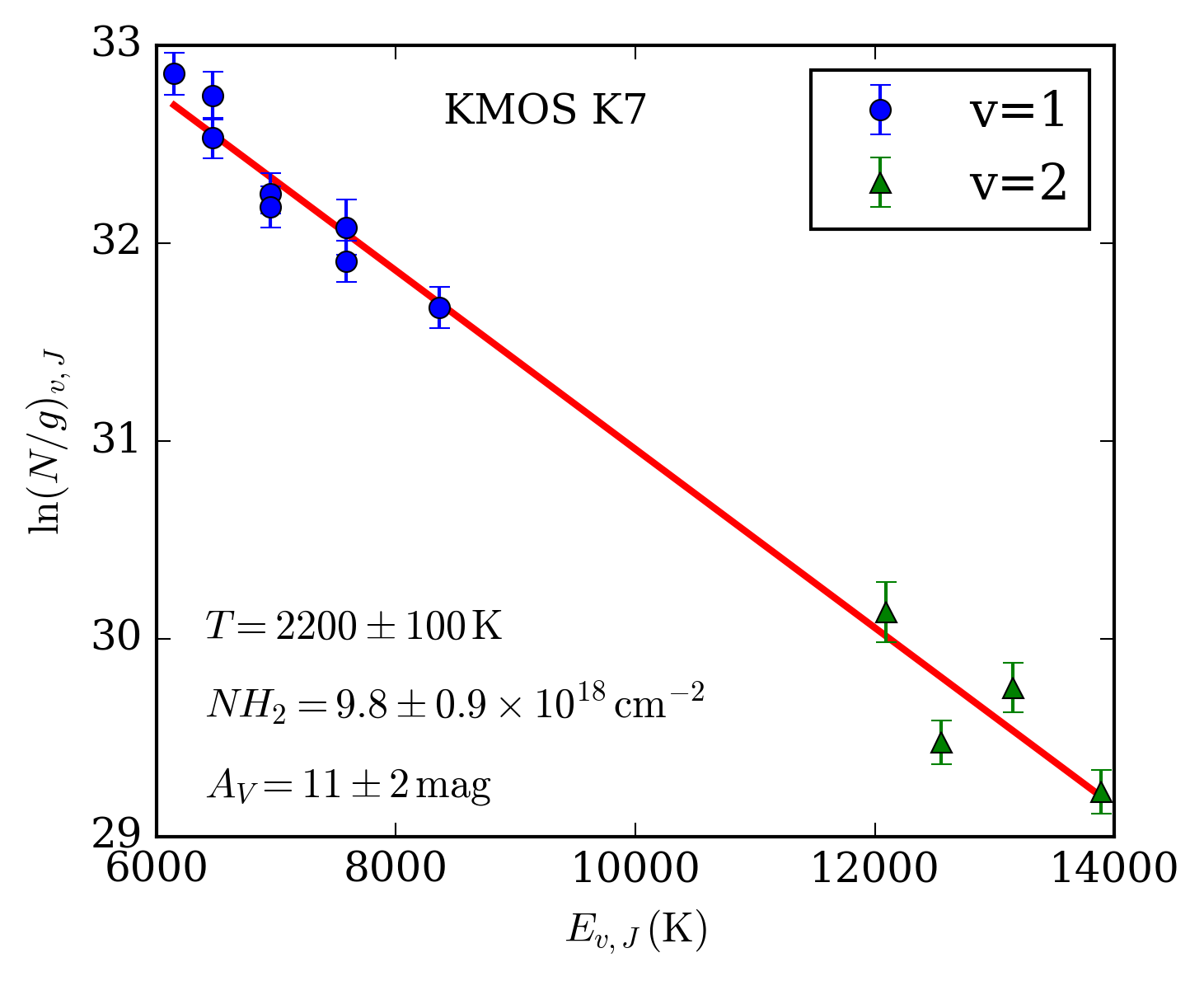}
    \includegraphics[width=0.32\hsize]{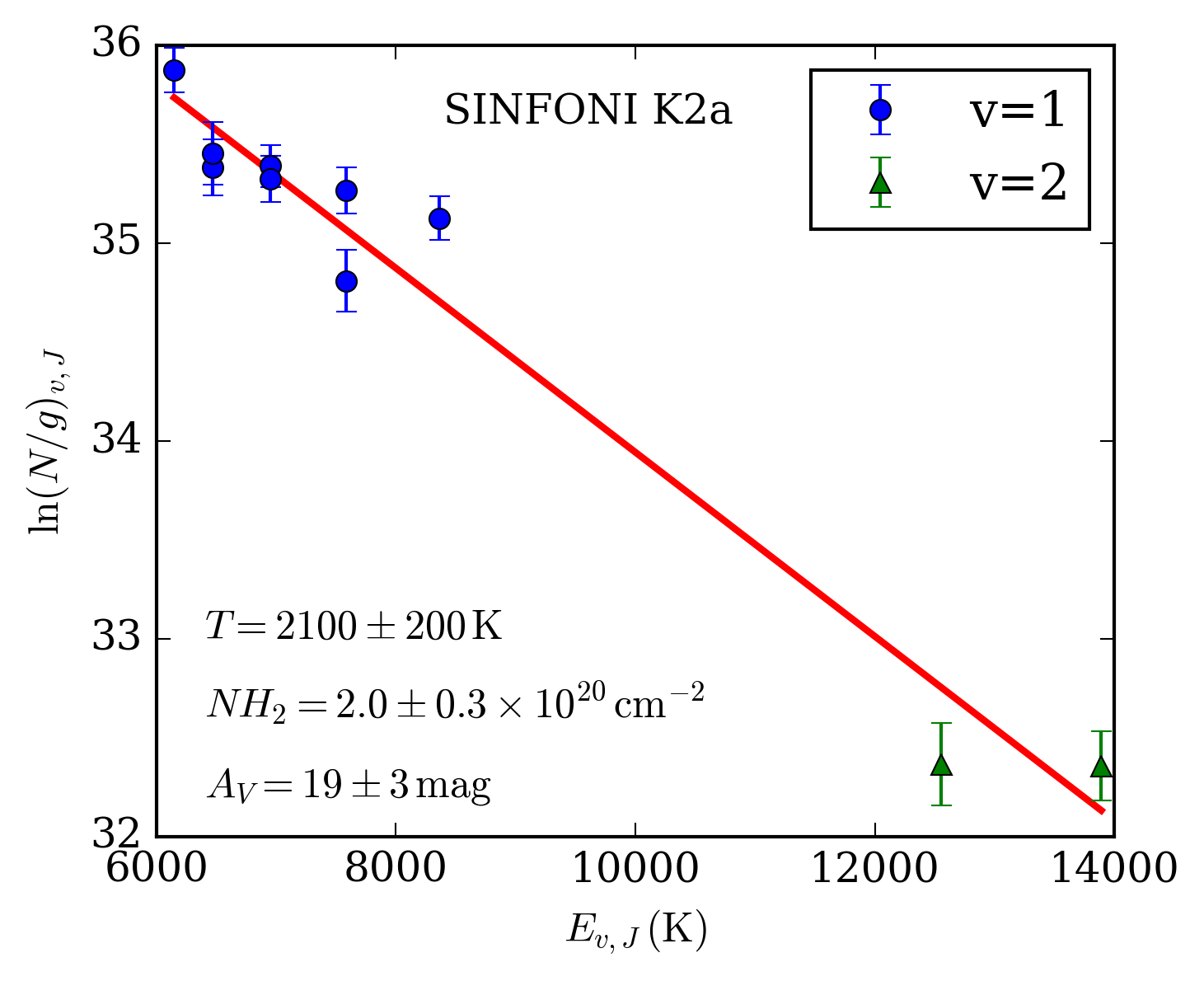}
    \includegraphics[width=0.32\hsize]{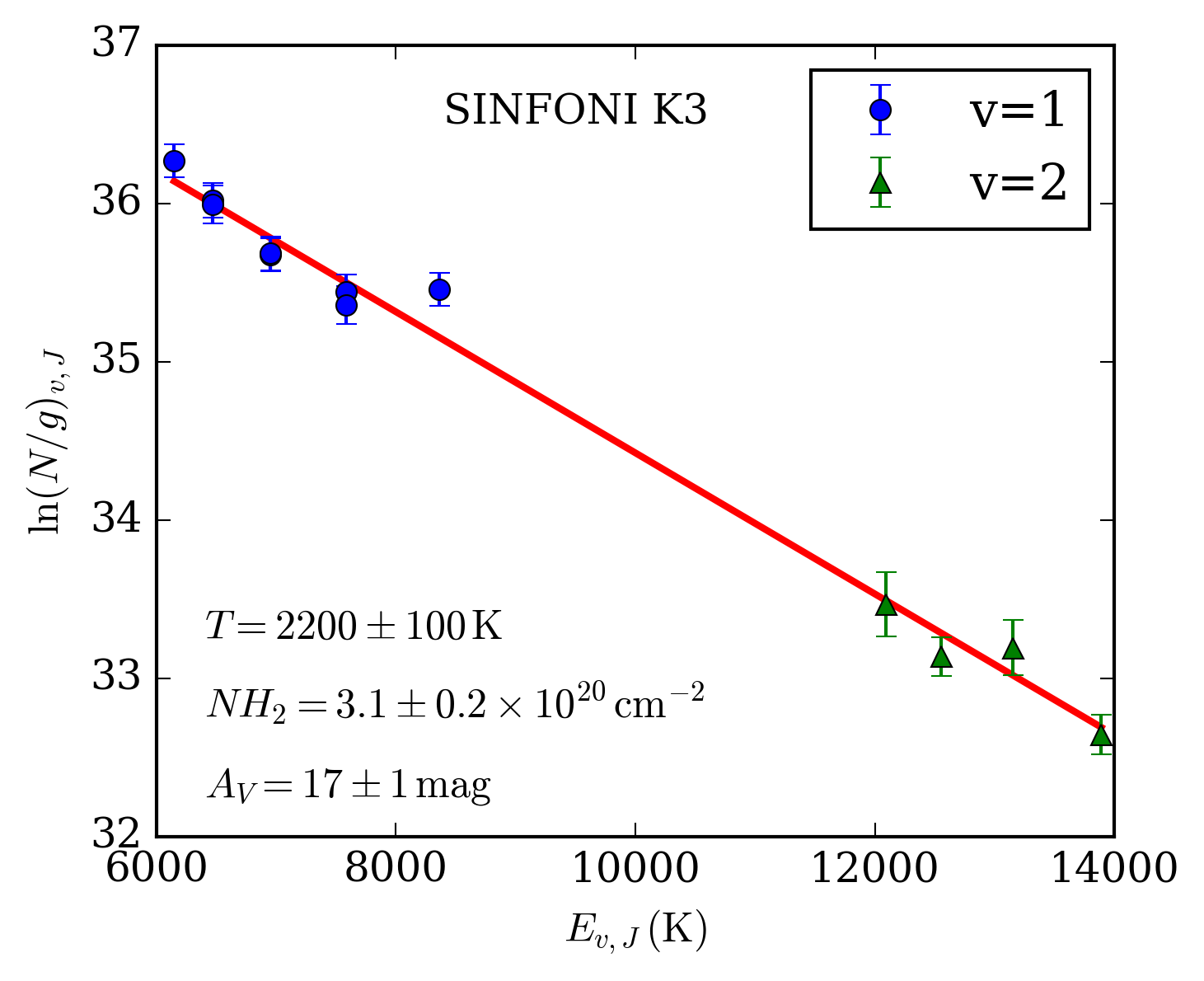}
    \includegraphics[width=0.32\hsize]{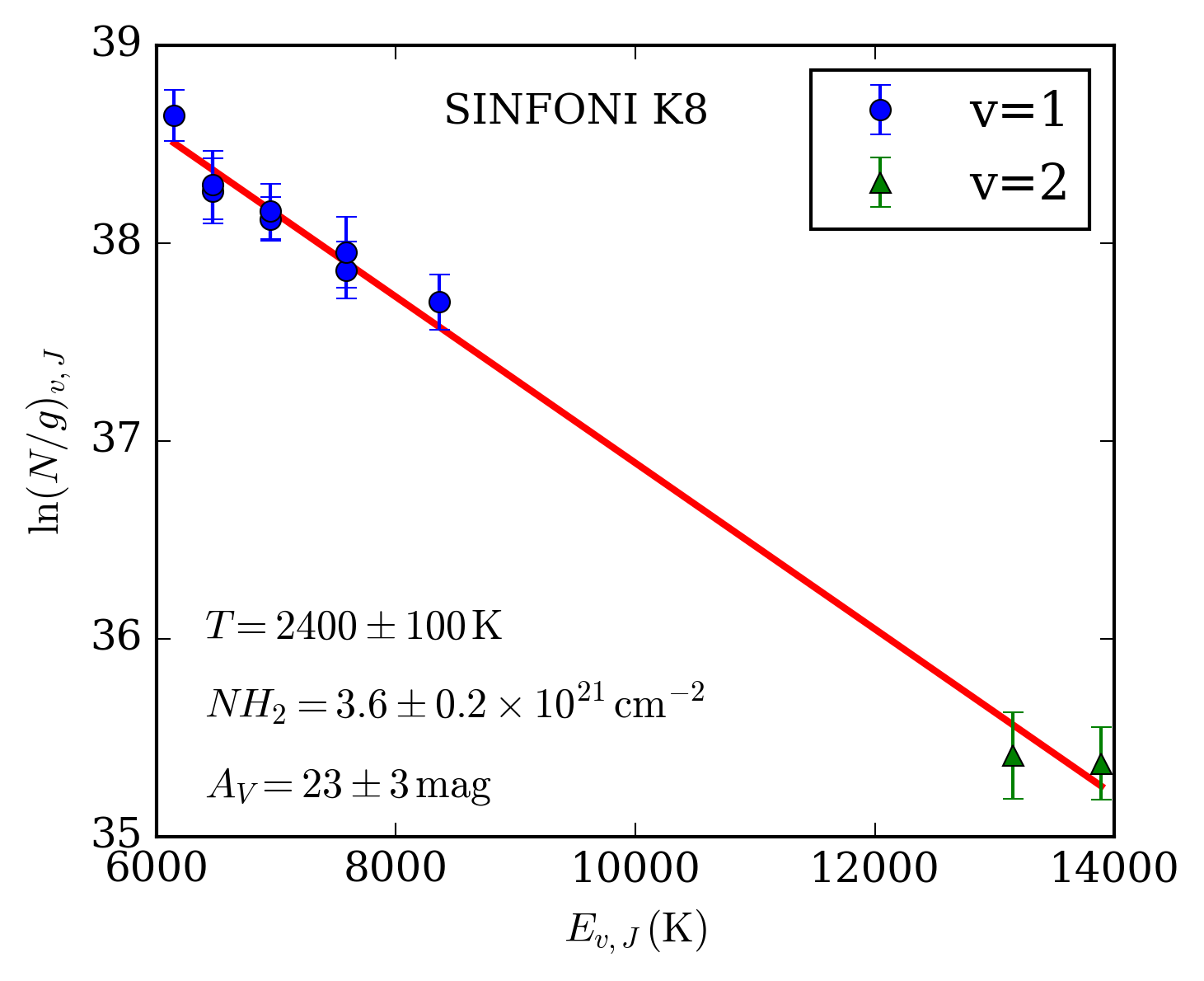}
    \caption{Ro-vibrations diagrams for all the \hmol\ knots outlined in the SINFONI and KMOS data sets. Labels identifying the knot and data set are at the top of each diagram. The KMOS knot K6 does not have a ro-vibrational plot as we could not derive a visual extinction value for it, and thus could not fit the plot to derive further physical parameters.}
    \label{fig:all_RV_plots}
\end{figure*}

Fig.\,\ref{fig:all_RV_plots} shows the ro-vibrational diagrams for the NIR knots presented in this work.

\section{Flux densities for SED}
\label{SED_appendix}

\begin{table}[h]
    \caption{Background subtracted flux densities derived at each wavelength for the SED plot.}
    \label{tab:SED_flux_densities}
    \centering
    \begin{tabular}{lc}
    \hline \hline
    \noalign{\smallskip}
    Wavelength & Flux density \\
    (\micron) & (Jy) \\
    \noalign{\smallskip}
    \hline
    \noalign{\smallskip}
    \textit{Spitzer} & \\
    3.6 & 0.12 (0.21)\\
    4.5 & 0.43  (0.52)\\
    5.8 & 0.84  (1.43)\\
    8.0 & 0.71  (2.59)\\
    24.0 & 4.76  (6.58)\\
    \noalign{\smallskip}
    \hline
    \noalign{\smallskip}
    \textit{WISE} & \\
    3.4 & 0.02   (0.08)\\
    4.6 & 0.50   (0.63)\\
    11.5 & 0.60   (4.21)\\
    22.0 & 4.64   (19.95)\\
    \noalign{\smallskip}
    \hline
    \noalign{\smallskip}
    \textit{Herschel} & \\
    70.0  & 310.48   (334.29)\\
    160.0 & 581.45   (660.17)\\
    350.0 & 89.16   (124.35)\\
    500.0 & 15.32   (32.90)\\
    \noalign{\smallskip}
    \hline
    \end{tabular}
    \tablefoot{The values in parenthesis are fluxes without the continuum subtraction. The photometric aperture radius was fixed at 16\arcsec, which was based on the $70\,\mu$m Herschel image \citep[see][for details in choosing the aperture size]{debuizer2017}.}
\end{table}

\begin{figure*}[h]
    \centering
    \includegraphics[width=1.0\hsize]{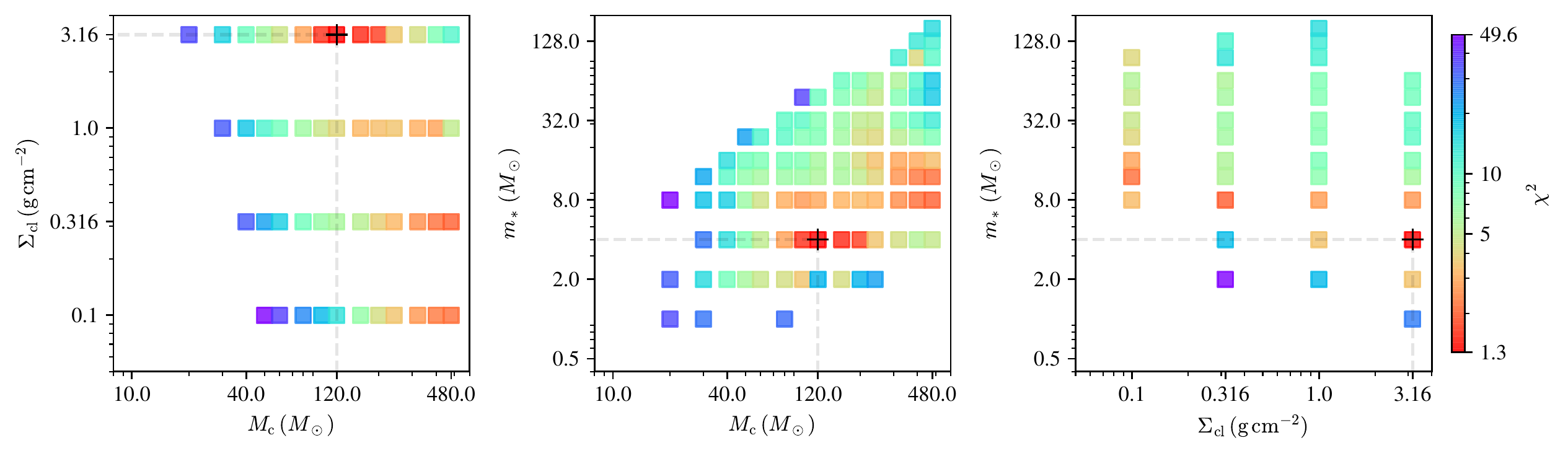}
    \caption{Diagrams of $\chi^2$ distribution in $\Sigma_\mathrm{cl}-M_\mathrm{c}$ space, $m_* - M_\mathrm{c}$ space, and $m_* - \Sigma_\mathrm{cl}$ space, where the black cross marks the location of the best model.}
    \label{fig:2D_plot}
\end{figure*}

\begin{table*}[h]
    \caption{Parameters of the five best fit models to the \iras\ when considering 50\% of the fluxes to construct the SED.}
    \label{tab:SED_parameters_50PERC}
    \centering
    \begin{tabular}{cccccccccccc}
        \hline \hline 
        \noalign{\smallskip}
        $\chi^2$ & $M_\mathrm{c}$ & $\Sigma$ & $R_\mathrm{core}$ & $m_*$ & $\theta_\mathrm{view}$ & $A_V$ & $M_\mathrm{env}$ & $\theta_\mathrm{w,esc}$ & $\dot{M}_\mathrm{disc}$ & $L_\mathrm{bol,iso}$ & $L_\mathrm{bol}$ \\
        $\mathrm{}$ & ($\mathrm{M_{\odot}}$) & ($\mathrm{g\,cm^{-2}}$) & ($\mathrm{pc}$) (\arcsec) & ($\mathrm{M_{\odot}}$) & ($\mathrm{{}^{\circ}}$) & ($\mathrm{mag}$) & ($\mathrm{M_{\odot}}$) & ($\mathrm{{}^{\circ}}$) & ($\mathrm{M_{\odot}\,yr^{-1}}$) & ($\mathrm{L_{\odot}}$) & ($\mathrm{L_{\odot}}$) \\
        \noalign{\smallskip}
        \hline
        \noalign{\smallskip}
            1.14 & 160 & 1.000 & 0.09 (5.6) & 4 & 13 & 169.33 & 150.68 & 10 & 2.6\ee{-4} & 1.3\ee{4} & 4.2\ee{3} \\
            1.16 & 200 & 1.000 & 0.10 (6.3) & 4 & 13 & 163.95 & 194.37 & 8  & 2.7\ee{-4} & 1.1\ee{4} & 4.7\ee{3} \\
            1.19 & 80  & 3.160 & 0.04 (2.5) & 2 & 13 & 206.56 & 77.00  & 9  & 3.6\ee{-4} & 1.6\ee{4} & 5.5\ee{3} \\
            1.25 & 60  & 3.160 & 0.03 (1.9) & 2 & 13 & 225.33 & 56.30  & 11 & 3.4\ee{-4} & 2.2\ee{4} & 5.4\ee{3} \\
            1.40 & 160 & 3.160 & 0.05 (3.1) & 2 & 13 & 113.93 & 157.54 & 5  & 4.4\ee{-4} & 8.5\ee{3} & 5.2\ee{3} \\
        \hline
    \end{tabular}
    \tablefoot{From left to right, the parameters are: $\chi^2$, initial core mass, mean mass surface density of the clump, initial core radius, current protostellar mass, viewing angle, foreground extinction, current envelope mass, half opening angle of outflow cavity, accretion rate from disc to protostar, isotropic bolometric luminosity, intrinsic bolometric luminosity.}
\end{table*}

\end{appendix}

\end{document}